\documentclass[journal=jacsat, manuscript=article]{achemso}
\usepackage{url}
\usepackage{tabularx,colortbl}
\usepackage{natbib}
\usepackage{mciteplus}
\usepackage[utf8]{inputenc}
\usepackage{hyperref}
\usepackage{hhline}
\usepackage{amsmath}
\usepackage{graphicx}
\usepackage{caption}
\usepackage{subcaption}
\usepackage[section]{placeins}
\usepackage{ifpdf}
\author{Nataliya Portman}
\email{nataliya.portman@uoit.ca}
\affiliation[University of Ontario Institute of Technology]{Department of Physics, University of Ontario Institute of Technology, Oshawa, Ontario, L1H 7K4 Canada}

\author{Isaac Tamblyn}
\email{isaac.tamblyn@nrc.ca}
\affiliation[University of Ontario Institute of Technology]{Department of Physics, University of Ontario Institute of Technology, Oshawa, Ontario, L1H 7K4 Canada}
\alsoaffiliation[National Research council of Canada]{National Research Council of Canada, 100 Sussex Drive, Ottawa, Ontario, K1A 0R6 Canada}
\title{Sampling algorithms for validation of supervised learning models for Ising-like systems.}


\begin{document}

\begin{abstract}
In this paper, we build and explore supervised learning models of ferromagnetic system behavior, using Monte-Carlo sampling of the spin configuration space generated by the 2D Ising model. Given the enormous size of the space of all possible Ising model realizations, the question arises as to how to choose a reasonable number of samples that will form physically meaningful and non-intersecting training and testing datasets. 
Here, we propose a sampling technique called ``ID-MH" that uses the Metropolis-Hastings algorithm creating Markov process across energy levels within the predefined configuration subspace. We show that application of this method retains phase transitions in both training and testing datasets and serves the purpose of validation of a machine learning algorithm. 
For larger lattice dimensions, ID-MH is not feasible as it requires knowledge of the complete configuration space. As such, we develop a new ``block-ID" sampling strategy: it decomposes the given structure into square blocks with lattice dimension \(N \leq 5\) and uses ID-MH sampling of candidate blocks. Further comparison of the performance of commonly used machine learning methods such as random forests, decision trees, k nearest neighbors and artificial neural networks shows that the PCA-based Decision Tree regressor is the most accurate predictor of magnetizations of the Ising model. For energies, however, the accuracy of prediction is not satisfactory, highlighting the need to consider more algorithmically complex methods (e.g., deep learning).     
\end{abstract}
\section*{Introduction}

Machine learning has been gaining attention in materials research and accelerated material discovery. Recently, machine learning approaches have shown the power to effectively learn from the past data and first principles materials characterization. Successful examples include fast prediction of phase diagrams \cite{art1} and crystal structures \cite{Morgan:2003}, approximation of density functionals \cite{Snyder:2012} and interatomic potentials \cite{Bartok:2010, *Behler:2011, *Morawietz:2013, *Botu2:2015, *Botu1:2015} for efficient and accurate materials simulation and prediction of physical properties from knowledge of similar known materials that avoids laborious computations \cite{Pilania:2013, Hansen:2013, Hansen:2015}. Most of these applications are based on \textit{supervised learning} methods. In a supervised learning framework, a mapping is learned between predictor and outcome variables describing a physical phenomenon under study in an algorithmic way. Such an algorithm is trained on a large collection of observations or measurements of predictors and their corresponding outcomes for the purpose of predicting the outcome from a new instance. For a continuous outcome, regression algorithms are used to predict electronic structure-property relationships (e.g., kernel ridge regression or regularized least squares model of Density Functional Theory Hamiltonian, neural networks for prediction of crystal structures \cite{Tatlier:2011, Rupp:2012, Schutt:2013, Hegde:2016, Brockherde:2016}). For a categorical outcome, classification algorithms are used to classify crystal structures or to detect material defects such as decision trees, support vector machines and neural networks \cite{Pilania:2015}.          

While machine learning has been making strides in condensed matter physics research, discovering phase transitions \cite{Wang:2016} and complex topological phases \cite{Carrasquilla:2016} of the Ising model-based ferromagnetic systems, other applications of supervised learning methods such as prediction of Ising system physical observables, have received little, if any, attention. In \cite{Bian:2010}, the authors used a quantum annealing hardware designed according to the Ising model with local neighbor interactions to find the global minimum and to sample Ising energy functions at zero and finite temperatures, respectively. This is an alternative to running software that realizes learning algorithms on classical computers.   

To apply supervised learning to 2D Ising ferromagnetic systems, we need to create a so-called training or representative dataset with various spin arrangements on a 2D lattice (configurations) as predictor variables and their corresponding total energies, for example, as outcome variables. Once the supervised learning model is trained on this dataset, we can test it on the other dataset. 

It seems straightforward to generate the representative dataset using Metropolis Monte-Carlo simulations leading to Boltzmann probability distribution of spin configurations over energy states. From a machine learning perspective, this approach suffers two major drawbacks:
\begin{enumerate}
\item Uneven Boltzmann distribution may lead to imbalance between energy classes in the training dataset resulting in a biased prediction of energy values.
\item Metropolis Monte-Carlo simulations collect repeated spin configurations that typically occur when the candidate configuration is rejected or proposed again. As a result, a traditional validation that involves repeated partitioning of the dataset may create intersecting testing and training subsets, yielding 100\% accuracy for ``shared" configurations and underestimating overall error.
\end{enumerate}
Thus, an unbiased sampling method is needed for training and testing data collection. This paper focuses first on the development of such sampling technique called ``block-ID'' (bID). Next, bID-sampled training and testing datasets are used with different machine learning algorithms to determine the best performing algorithm for the specific case of 2D Ising model of the ferromagnetic system. 

More precisely, we address the question of machine learning of major physical quantities described by the Hamiltonian \(\mathcal{H}\) and Magnetization \(\mathcal{M}\) operators as follows 
\begin{align}
\label{eq:ref1}
\mathcal{H} & =-J \sum_{(i,j)} \sigma_i \sigma_j-h\sum_{i=1}^{N^2} \sigma_i ,\\
\mathcal{M} & =\sum_{i}^{N^2} \sigma_i .
\end{align}
Here \( \sigma_i \) are random spin variables assuming the values \( \pm{1} \) on the sites of a 2D square lattice of size \( N \times N\) directed either up or down.  
The first term in (\ref{eq:ref1}) is the sum of the nearest neighbor site interaction energies. The second term describes the interaction of the applied magnetic field \( h \) with the spin system. Without loss of generality, we assume no external field. The proposed validation method developed in this paper consists of two steps: 
\begin{enumerate}
\item Generation of training and testing datasets using bID sampling algorithm,
\item Statistical measurement of the machine learning algorithm performance (via median error). 
\end{enumerate}
It is applicable to cases with a non-zero external field. Python code realizing the proposed sampling technique is included as a supplement to this paper.

Overall, this work enables further exploration and exploitation of supervised learning techniques for other Ising model-based distinct physical systems than ferromagnetic ones. Examples of such systems include binary alloys where the spin variables represent atoms of type \( A \) or \( B \) and lattice gas where \(+1\)/\(-1\) indicate the presence/absence of a molecule. We believe that with the proposed validation method, new accurate and physically informed predictive models of energy or other quantities of interest can be discovered and applied to computationally challenging cases of molecular lattice-like structures on a larger length scale (e.g., crystalline materials). For example, block-ID sampling algorithm allows to build training and testing data models of binary aluminum-based systems with dopant elements (e.g., titanium, silver, etc.) whose atoms are arranged in a 3D cubic lattice that is a unit cell of size \(N=1000\) or more.               
\section*{Paper organization}

This paper is organized as follows. First, we show the motivation for a novel sampling technique from a machine learning perspective, and introduce an ID-MH (ID-Metropolis-Hastings) algorithm for the Ising model of an \(N \times N\) lattice for \(N \leq 5\) (Sections \textbf{\nameref{sec1}} and \textbf{\nameref{sec2}}). We then incorporate the ID-MH method into a so called \textit{block-ID} (bID) sampling scheme for the Ising model on a larger length scale equal to a multiple of \(N\) (Section \textbf{\nameref{sec3}}). We further investigate the structure of the datasets generated by bID sampling using PCA-based dimensionality reduction and the qualitative behavior of the ferromagnetic system (Sections
\\
 \textbf{\nameref{sec4}}
\\
and \textbf{\nameref{sec5}}). 
\\
Section \textbf{\nameref{sec6}} compares the performance of common machine learning approaches, namely, \textit{k nearest neighbors, decision trees, random forests, artificial neural networks} trained on raw Ising configurations (collected by bID sampling) and on PCA-based features of the configuration set. Also, we study the effect of energy distribution equalization on the performance of the algorithms. Section \textbf{\nameref{sec7}} concludes the work and outlines future directions. 

\section{Motivation}
\label{sec1}
Metropolis Monte-Carlo (MMC) \cite{Newman:1999} sampling of representative configurations/microstates from high-dimensional probability distributions provides the basis for calculation of equillibrium properties of Ising model-based systems (e.g., internal energy, net magnetization). The efficiency of MMC method lies in sampling a small but important fraction of the system microstate space, resulting in a narrow range of energies and other physically relevant quantities. Real systems connected to a thermal bath, will, at equillibrium, sample the microstates according to the Boltzmann probability distribution (\ref{boltz})  that determines a small number of dominating energy states,
\begin{equation}
\begin{centering}
\label{boltz}
p(C)=\frac{1}{Z}\exp{\left(-\frac{E(C)}{k_B T}\right)}.
\end{centering}
\end{equation}
Here \(E(C)\) is the energy of microstate \(C\), \(k_B\) is Boltzmann constant and Z is the partition function. For simplicity of notation, we use \(\beta=\frac{1}{k_B T}\) in the definition of Boltzmann weights given by (\ref{boltz}).

MMC has been an algorithm of choice for the study of equillibrium properties of many-particle systems for the past 50 years. Compared to other Monte-Carlo variants of Markov Chain type, the MMC algorithm samples Boltzmann distributions more efficiently due to its rapid convergence to equillibrium. Microstates in the asymptotic part of the Markov chain generated by Metropolis MC sampling form a canonical ensemble from which all macroscopic properties of the system can be calculated.

In \cite{Carrasquilla:2016} the authors created such canonical ensembles for different temperatures combining them into the training dataset for the purpose of machine learning phase transitions of the classical Ising model. They demonstrated that Convolutional Neural Networks trained on this dataset are capable of identifying various phase transitions including non-trivial Coulomb and topological phases. The same canonical ensembles were also used in an unsupervised setting to discover distinct phases and their salient features, showing the power of learning techniques in the study of equillibrium behaviour of condensed matter systems \cite{Wang:2016}.

The success of machine learning implementations depends on the ``right" training dataset - it should be physically informed by the phenomenon under study. Carrasquilla et al. were able to identify the ferromagnetic phase transition since it was implicitly built into the temperature-dependent dataset of Ising configurations (canonical ensembles). In essence, this training dataset represents the data model of ferromagnetic system behavior. 

Using a supervised learning approach, can we accurately predict equillibrium macroscopic properties of the Ising model at any given temperature? We envision building such a predictor with the application of Metropolis MC sampling of the space of Ising configurations/microstates where the energy of each newly accepted configuration \(E(C_i)=\mathcal{H}(C_i)\) is forecasted via a machine learned Hamiltonian operator (\ref{eq:ref1}). Since Metropolis MC samples Ising configurations \(C_i\) with Boltzmann weights \(\frac{\exp{\left( -\beta E(C_i) \right)}}{Z}\), a physical observable of interest \( \langle Q \rangle \) is estimated as a simple arithmetic mean over the equillibrium ensemble \({C_1, C_2,...,C_M}\) 
\begin{equation}
\label{mean}
\langle Q \rangle \approx \frac{1}{M}\sum_{i=1}^{M} Q(C_i). 
\end{equation}
In this approach, the accuracy of prediction of physical observables is greatly influenced by the accuracy of prediction of Hamiltonian energies as they determine Boltzmann probabilities of selecting microstates (\ref{boltz}). Therefore, a highly accurate, machine learned Hamiltonian operator will enable prediction of a variety of thermodynamic and/or ferromagnetic observables (magnetic susceptibility) depending on the system under study. 

In this paper, we treat the Hamiltonian and Magnetization operators (\ref{eq:ref1}) in a supervised learning paradigm, viewing them as mappings between an Ising configuration and its total energy and magnetization, respectively.           
We consider the Ising model with periodic boundary conditions where spins directed upward or downward are placed on sites of a 2D square lattice. 

With the task of supervised learning of microstate energies and magnetizations, inevitably comes another task of algorithm validation. As a common practice, a statistical method known as k-fold cross-validation has been used for the assessment of algorithm performance. It repeatedly partitions the sample of observed data into training and testing (hold-off) subsets containing at least \( 50\% \) of examples and at most \(50 \%\) of examples, respectively. At each partition, these examples (for the Ising model, they are combinations of spins arranged in a square lattice order) are sampled at random, and new training and testing datasets are created. The error of prediction (the absolute difference between predicted and true values) is then computed for each testing example and averaged over all testing data. 

For a MMC-generated dataset comprised of approximately canonical ensembles for different temperatures, k-fold cross-validation is not adequate for the following reasons:
\begin{enumerate} 
\item At low temperatures, the Ising spin system spends the majority of its time at the lowest-energy state represented by two distinct configurations. This is reflected in a large number of repeated configurations representing the lowest-energy state. As a result, there is a high chance of the presence of these configurations appearing in both training and testing datasets, leading to a significant overall error underestimate (given the dominant contribution of lowest energy testing samples).
\item Other repeated configurations naturally occur when new candidates are rejected at each MMC step (acceptance of a new candidate with probability) when the Markov chain has reached asymptotic equillibrium. Repeated configurations may appear in training and testing datasets by means of random sampling, resulting in a further underestimate of the overall error in energy/magnetization prediction.
\end{enumerate}
Another issue with a MMC-generated dataset is the imbalanced distribution of microstate representatives of energy states, yielding a bias in error estimates toward dominant, frequently visited energy states.

While MMC generates physically meaningful datasets (capturing ferromagnetic phase transition phenomenon and important energy ranges for different temperatures), its sampling bias will render prediction of underrepresented energies inaccurate. From a machine learning perspective, more variability should be built into these energy states in order to lessen the imbalance effect. 
We set out to devise a sampling algorithm that satisfies the following requirements:   
\begin{enumerate} 
  \item Physically motivated: 
  \begin{enumerate}
    \item Qualitatively captures qualitatively the equillibrium behaviour of the Ising model,
    \item Samples a relevant energy range.
  \end{enumerate}
  \item Generates non-intersecting and consistent training and testing datasets of Ising spin configurations, where consistency means inclusion of all energy states of the testing dataset into the energy range of the training dataset, and similarity of the probability distributions of energies and magnetizations.
  \item Oversamples underrepresented energy states obtained by Metropolis-MC.
  \item Applies to the Ising Model on a 2D lattice of any dimension \( N \).
\end{enumerate}
We propose ID-MH and block-ID sampling algorithms that satisfy the above mentioned requirements for \(N \leq 5\) and multiples of \(N\), respectively. 

\section{ID-MH sampling algorithm}
\label{sec2}
\begin{figure}
    \begin{center}
    \begin{subfigure}{.45\textwidth}
        \includegraphics[width=\textwidth]{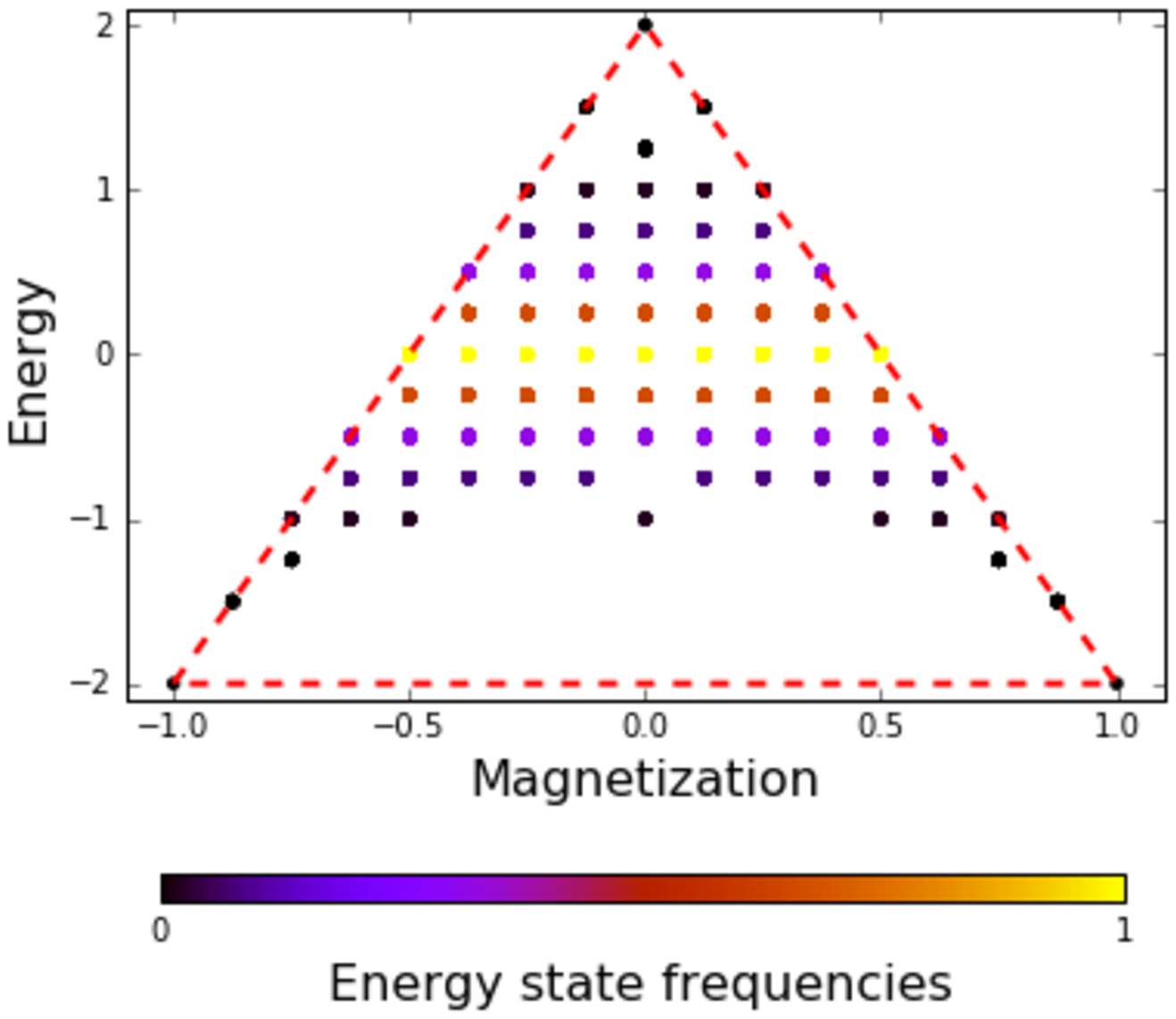}
        \centerline{(a)}
    \end{subfigure}
    \begin{subfigure}{.45\textwidth}
        \includegraphics[width=\textwidth]{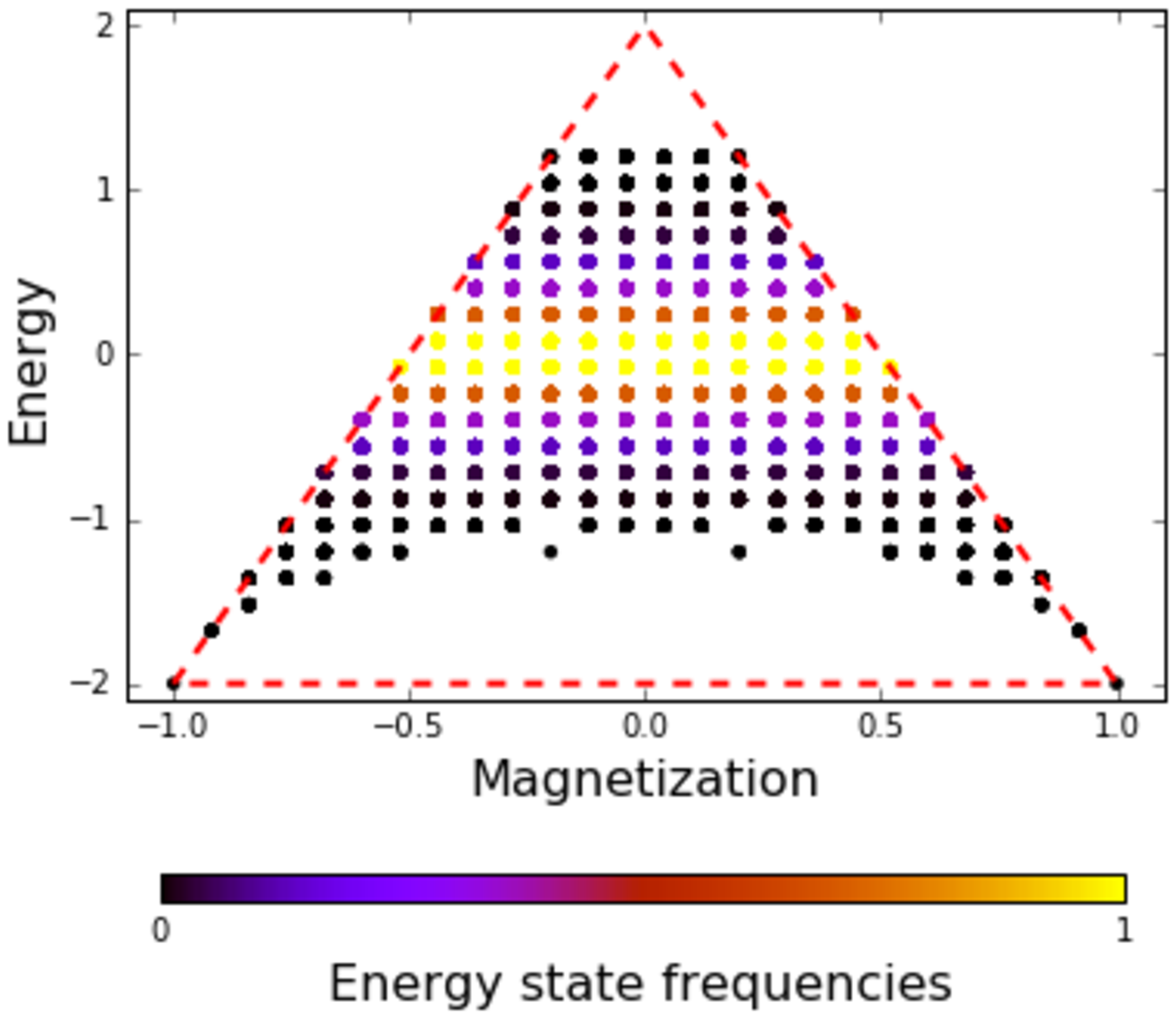}
        \centerline{(b)}
    \end{subfigure}
    \end{center}
\caption{``Phase space" plots of (a) the complete dataset of microstates for \(4 \times 4\) Ising model and (b) 30 \% of the complete dataset for \(5 \times 5\) Ising model. All microstates defined by energies and magnetizations are enclosed in a triangle with vertices whose y- coordinates correspond to lowest and highest energy values.}
\label{fig3a}
\end{figure}

We start with the consideration of the Ising model with periodic boundary conditions on a 2D square lattice of size \(N^2\), where \(N\) is a small number and with a coupling coefficient \(J=1\) for all pairs of interacting neighbors. The space of all possible combinations of spins, (\(1\) for upward direction and \(-1\) for downward direction) contains \(2^{N^2}\) configurations and increases exponentially with respect to \(N\). For \(N=4\), our system has 65,536 unique configurations (see Figure \ref{fig3a}.a), and for \(N=5\) the number of configurations increases to 33,554,432 (see Figure \ref{fig3a}.b). Figure \ref{fig3a} shows that all possible microstates defined by energies and magnetizations per site are enclosed by a triangle with vertices corresponding to lowest and highest energies per site. For each energy state, we counted the number of representative configurations and then  normalized all obtained frequencies to the range \([0.0, 1.0]\). Color-coded plots of energy frequencies in Figure \ref{fig3a} demonstrate symmetric distribution of microstates with respect to energy peaked at the \(0th\) energy level and evolution of magnetization into two branches as energy decreases.  

For \(N=6\) it would take \(\frac{2^{36}}{2^{25}}=2048\) times longer than the previous calculation to generate the complete dataset. Therefore, in order to avoid spending a long time on computer calculations of this kind, we limit ourselves to the case of the lattice dimension \(N \leq 5\). 
In nature, we do not encounter such small size Ising systems. However, if we find a clever strategy for creating consistent and non-intersecting training and testing datasets for small lattice sizes we can make it applicable to larger ones. 

A straightforward approach is to partition the complete Ising dataset by random drawing of its configurations and setting them aside for testing. However, this approach does not scale linearly with the increase of \(N\). Such an approach requires generation of the complete dataset, which is impossible to do in a reasonable amount of time even for moderate values of \(N\). Moreover, the complete set of configurations is redundant. As mentioned earlier, for the purpose of prediction of equillibrium behavior of the Ising system, we are interested only in those configurations/microstates that the system is likely to visit at thermal equillibrium. Such microstates constitute a small fraction of the entire dataset.

Leaving scalability issue aside, we focus on the development of the sampling algorithm for small lattice dimension \(N \leq 5\) that will create non-intersecting and consistent training and testing datasets of microstates representing ``important" energy states. To fix ideas, let \(N=4\). Denote by \(\chi\) a finite set of all energy states. Then \( \chi \) contains the total of 15 energy values
\begin{equation}
\chi=\{-32,-24, -20,-16,-12,-8,-4, 0, 4, 8, 12, 16, 20, 24, 32\}
\end{equation}
 with the frequencies recorded in a Table \ref{tab1} below.
 \newline
 \begin{table}[!h]
 \small
\begin{tabularx}{1.2\textwidth}{ |c|c|c|c|c|c|c|c|c| }
 \hhline{---------}
 Energy &-32 & -24 & -20 & -16 & -12 & -8 & -4 & 0  \\
 \hhline{---------}
 Counts & 2    & 32 & 64 & 424    & 1728 & 6688 & 13568 & 20524 \\
 Freq. & 3.05e-05 & 4.88e-04 & 9.76e-04 & 6.47e-03 & 2.63e-02 & 1.02e-01    & 2.07e-01 & 3.13e-01  \\
\hhline{---------}
\end{tabularx}
\vskip 3mm
\begin{tabularx}{1.2\textwidth}{ |c|c|c|c|c|c|c|c| }
 \hhline{--------}
 Energy & 4 & 8 & 12 & 16 & 20 & 24 & 32 \\
 \hhline{--------}
 Counts & 13568 & 6688 & 1728 & 424 & 64 & 32 & 2\\
 Freq. & 2.07e-01 & 1.02e-01 & 2.63e-02 & 6.46e-03 & 9.76e-04 & 4.88e-04 & 3.05e-05 \\
\hhline{--------}
\end{tabularx}
\caption{Frequencies of energy states in the complete space of \(4 \times 4 \) Ising configurations. }
\label{tab1}
\end{table}

\begin{figure}[ht]
\centering
\includegraphics[width=0.8\linewidth]{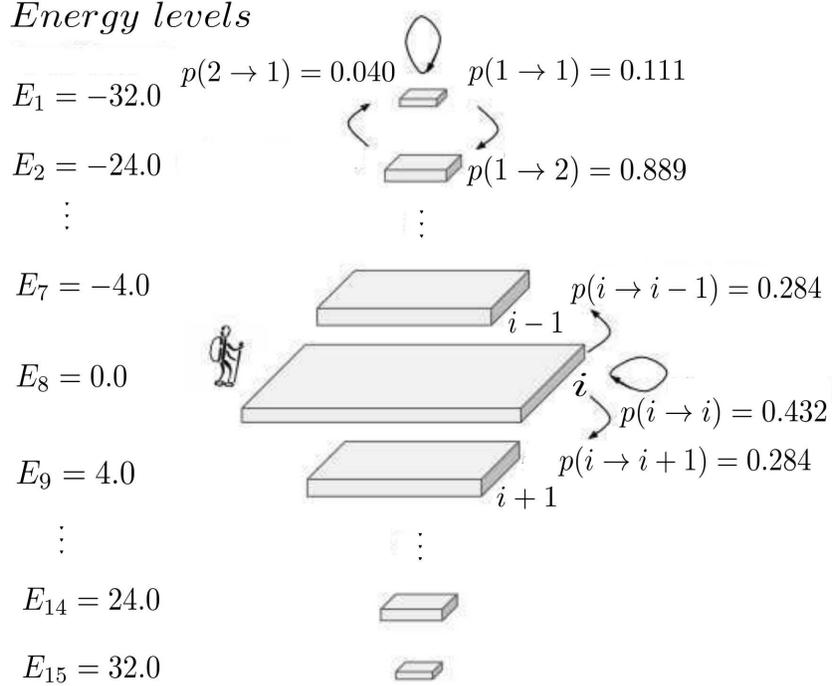}
\caption{ID-MH sampling algorithm for the \(4 \times 4\) Ising model. Microstates are sampled through the Markov Chain Monte-Carlo process formed across the energy ``decks" of the complete dataset of Ising configurations. Transition probabilities are defined from an  \textit{a priori} probability distribution of energy states.}
\label{fig4}
\end{figure}

Table \ref{tab1} shows the number of microstates having the same energy value \( E_i \in \chi \) for \( i \in \mathcal{N}: 1 \leq i \leq 15 \) and forming degenerate energy ``decks". For the simplicity of notation, we enumerate the decks in the ascending order of their energy values from 1 to 15 (see Figure \ref{fig4}). To devise a sampling algorithm in Markov Chain Monte-Carlo fashion, we have to specify transition and stationary probabilities of finding a configuration with energy \(E_i\) \cite{Krauth:1996}. Denote by \(p (i \rightarrow j)\) the probability of moving from deck \(i\) to deck \(j\). Setting up the transfer matrix of the Markov Chain Monte-Carlo algorithm such that 
\begin{equation} 
p(i \rightarrow i+1) + p( i \rightarrow i) +p(i \rightarrow i-1)=1 \text{ for } i \in \mathcal{N}: 2 \leq i \leq 14, 
\end{equation}
and for boundary decks
\begin{equation} 
p( i \rightarrow i) +p(i \rightarrow i \pm 1 )=1 \text{ for } i \in \{1, 15 \} 
\end{equation} 
hold in matrix columns \(i\), we restrict the moves from the current deck \(i\) to the neighboring decks \(i-1\) and/or \(i+1\) only. We define these probabilities using \textit{a priori} weights of energy decks recorded in the frequency row of Table \ref{tab1}. For example, if walkers are located on deck \(i\) with a larger number of representatives than those of the neighboring decks, then they are most likely to stay there and pick a candidate configuration from deck \( i \) (see Figure \ref{fig4}). If walkers are on the deck with the lowest energy value, then they are most likely to move to the deck with the second lowest value (as it has 16 times more microstates to choose from).

Let us choose deck \(i=8\) corresponding to \(0th\) energy value. Given \textit{a priori} probabilities of energy states \(7, 8, 9\) in the complete space of Ising configurations equal to \(0.207, 0.313, 0.207\), respectively, we normalize them to the unit sum and define the transition probabilities as
\begin{align}
p(8 \rightarrow 7) = 0.284\\
p(8 \rightarrow 8) = 0.432\\
p(8 \rightarrow 9) = 0.284
\end{align}
Here, the transition probabilities of moving in the opposite direction, that is, from \(9 \rightarrow 8 \) and from \(7 \rightarrow 8\), are not equal, leading to Metropolis-Hastings Monte-Carlo algorithm formulation.
While transition probabilities allow walkers to explore a vast variety of microstates, we further restrict the walker moves such that, after a large number of steps, they collect these configurations with Boltzmann probabilities. We define the stationary probability of selecting a configuration \(C_k\) from \(jth\) deck as 
\begin{equation}
\pi(C_k^j)=\frac{\exp{(-\beta E_j)}}{Z}.
\end{equation}
We have arrived at the following algorithm that produces a Markov chain of random energy states on \(\chi\), collects their representative microstates, and converges to the distribution \(\pi\):
\begin{enumerate}
\item Generate a random Ising configuration \(C_l^i\) and let its energy \(E_i\) be an initial state of the Markov chain.
\item Generate a candidate energy deck \(j\) from the distribution \(p(\cdot | i)\), specified by the frequencies of energy states in the complete dataset of spin combinations.
\item Draw a random configuration \(C_k^j\) from the candidate deck.
\item Calculate the acceptance ratio \(A=min(1, exp{(-\beta (E_j-E_i))}\frac{p(i|j)}{p(j|i)})\).
\item Generate a random variable \( u ~ U(0,1) \).
\item If \(u < A\) or \(E_j-E_i<0\), then accept \(C_k^j\), otherwise stay at current configuration of energy deck \(i\).
\item Repeat the process from step 2 to 6 until convergence and record last 1000 or more configurations, labeled by their energies and magnetizations.
\end{enumerate}
The difference between Metropolis-MC and Metropolis-Hastings-MC sampling lies in the choice of a candidate configuration. In Metropolis-Hastings, we have incorporated \textit{a priori} probability distribution of all energy states that defines the ``proposal" probability \(p(\cdot |i)\) on \(\chi\) from which the candidate configuration is drawn. In Metropolis, a candidate configuration is formed by a flip of a single spin that shifts the current energy state by 4 units up or down or does not change the energy state at all. This is equivalent to its drawing from a neighboring energy deck located below or above the current one or staying at the same energy deck. No preference is given for the choice of the neighboring deck, and the``proposal" probability \(p(\cdot |i)\) is symmetric. 
We call the proposed Metropolis-Hastings (MH) algorithm an \textit{ID-MH} algorithm as it samples the space of unique Ising configurations that can be enumerated and labeled by ID numbers.  

For the purpose of generating non-intersecting datasets, we further restrict the ID-MH Markov process to sample a certain subspace of Ising configurations. Indeed, if we sample at random 50\% of all Ising configurations and form two subspaces out of the sampled data and the remaining 50\%, then running ID-MH algorithms on each of these subspaces will yield two non-intersecting datasets. However, the way how we separate into two subspaces is important to the success of machine learning algorithms. From Table \ref{tab1}, we observe that ``tail", or boundary energy states contain only two representative configurations each. The lowest energy configurations contain all \(1s\) or \(-1s\), and the highest energy ones contain \(1s\) and \(-1s\) arranged in a checkered pattern. This holds true for all lattice dimension values of \(N\). If the training dataset contains only one example in the boundary energy states, then another representative example will not be recognized (by a trained predictive model) to have the same lowest/highest energy value due to its dissimilarity to the training example (having all spins in the opposite direction). Therefore, as a first step in forming non-intersecting and consistent datasets, we place the above mentioned representatives of boundary energy states to the training dataset. 

Next, it is important to realize that a simple split of the Ising dataset (without boundary classes) by means of random drawing of 50 \% of its microstates may lead to a significant imbalance in numbers of representatives of the same energy class in testing and training datasets. For example, if the number of configurations in some energy state in the training dataset turns out to be small, then a trained predictive model will not generalize well to other configurations found in that energy state. In order to avoid this issue, we take more control over random sampling procedure for splitting the complete space of Ising configurations into two subspaces. Specifically, we perform the following steps:
\begin {enumerate}
\item  Distribute Ising configurations to their corresponding energy decks (keeping in mind that we have already placed boundary decks to the training dataset).
\item In each energy deck:
    \begin{enumerate}
    \item shuffle its representatives,
    \item split them into 50 \% of training and 50\% of testing configurations using random sampling.
    \end{enumerate}
\end{enumerate}    
We call the datasets obtained by this splitting process \textit{candidate training/testing subspaces}.
\begin{figure}[!h]
\centering
\includegraphics[width=0.8\linewidth, height=490pt]{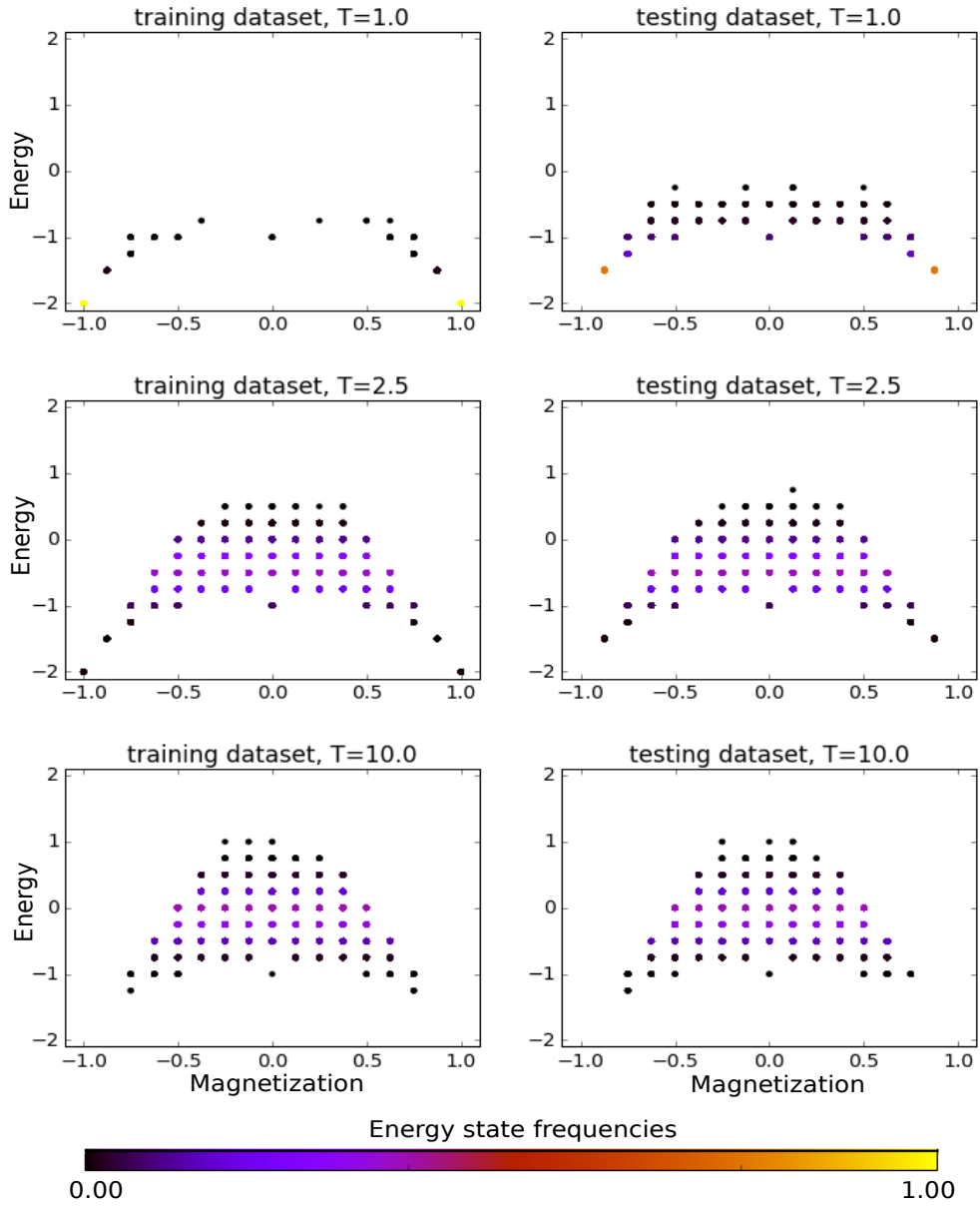}
\caption{``Phase space" plots of training and testing datasets for the \(4 \times 4\) Ising model. Higher energy states that are not present in the training dataset tend to appear in the testing dataset as temperature decreases. }
\label{fig5}
\end{figure}

In this way, we ensure that training and testing data models built by ID-MH sampling of candidate training and testing subspaces equally capture the variability of microstates representing important energy range. Also, data consistency is maintained, that is, the energy range of the candidate testing subspace is included into that of the candidate training subspace. We ensure that all energy states captured by the testing data model are present in the training data model.

Having created candidate training and testing Ising subspaces, we now run ID-MH algorithm for 40 initial random configurations that we refer to as \textit{seeds} at temperatures \(\frac{1}{\beta}=1.0, 2.5\) and \(10.0\). The reason for generating 40 seed trajectories through each of the candidate subspaces is to avoid falling into the local minimum of the global energy of a spin configuration when converging to the equillibrium distribution. Having recorded last 1000 microstates of each seed trajectory consisting of 32,000 MC steps across the energy decks of the candidate subspace, we arrive at the similar probability distributions of energy- and magnetizaton- dependent states in training and testing datasets (see Figure \ref{fig5}). At low temperature \(\frac{1}{\beta}=1.0\) the ferromagnetic systems mostly populate magnetization per site branches with values \(-1\) and \(1\) in the training dataset. Same behavior is observed in the testing dataset where these systems tend to visit the lowest energy state (which is equal to the second lowest value of the energy in the complete dataset of Ising microstates) present in this dataset. As the temperature increases to \(\frac{1}{\beta}=2.5\), the probability peak moves closer to the \(0th\) value of the energy per site in both datasets with rare occurrences of spontaneous magnetization. At a higher temperature, the system has fully lost its spontaneous net magnetization and the frequency peak gets established at the \(0th\) energy state giving the preference to chaotic or unstructured combinations of spins.  
\begin{figure}[ht]
\begin{center}
\includegraphics[width=.48\textwidth, height=140pt]{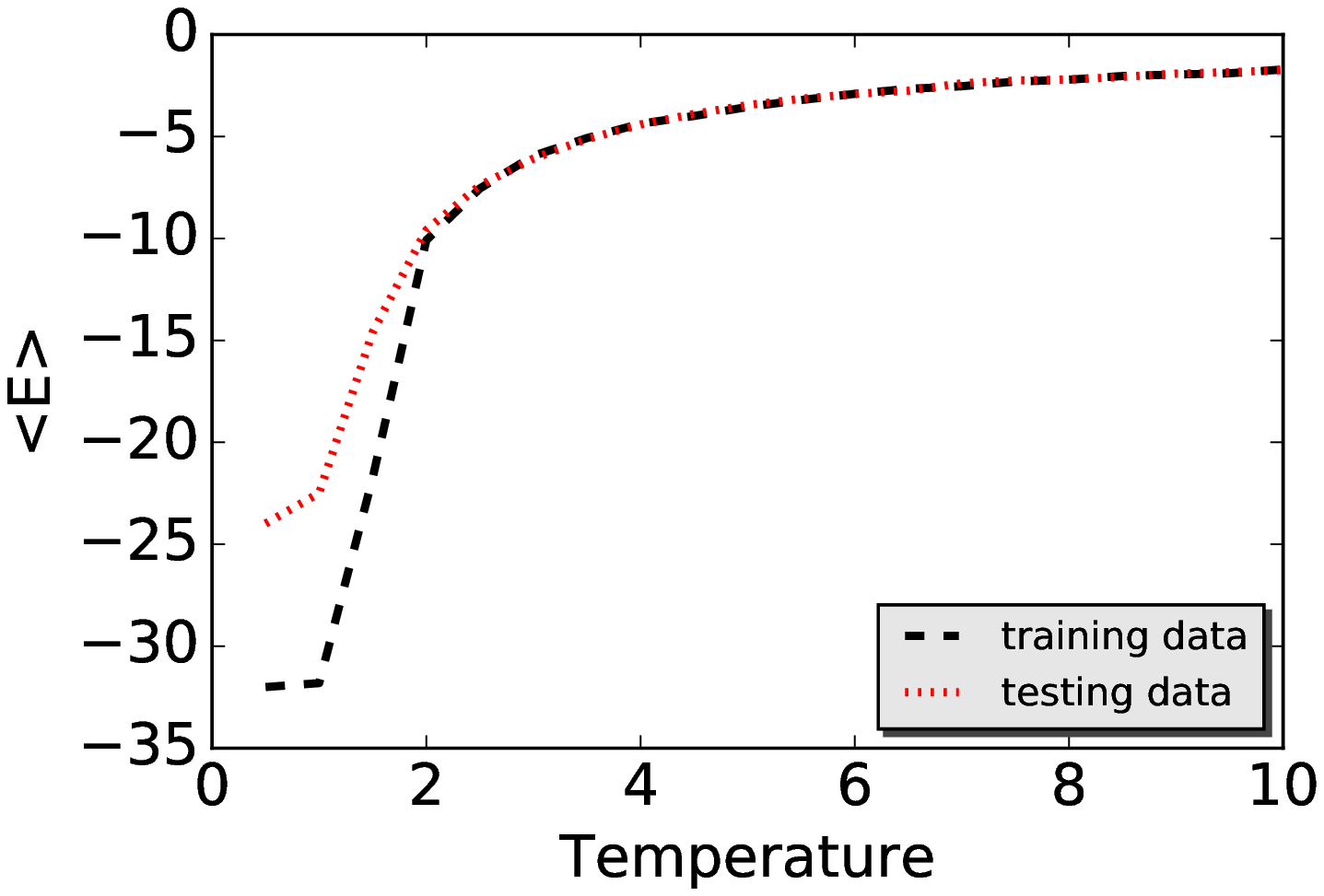}
\includegraphics[width=.48\textwidth, height=140pt]{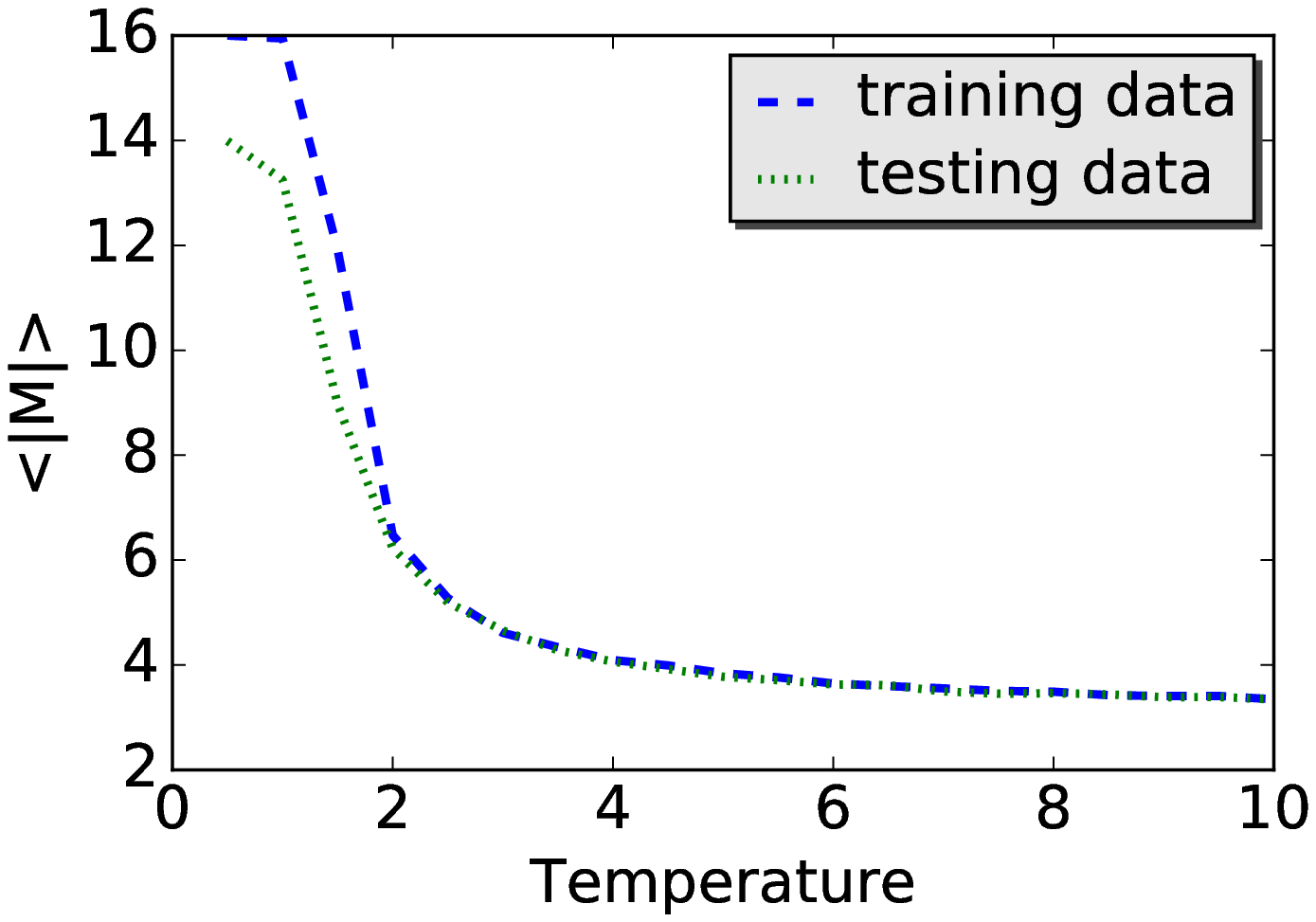}\\
\centerline{(a) \hspace{170pt} (b)}
\caption{Temperature-dependent, average (a) energy and (b) magnetization curves obtained by ID-MH sampling of \(4 \times 4\) Ising configurations. The discrepancy in average energies and magnetizations per site in training and testing datasets is due to the exclusion of the the lowest energy state from the candidate testing subspace.}
\label{fig6}
\end{center}
\end{figure}

We also observe the phenomenon of the phase transition in a rapid drop of mean absolute value of magnetization per site in its temperature-dependent curves in the vicinity of T=2.0 (see Figure \ref{fig6}.b). Simultaneously, the mean energy per site rapidly increases near T=2.0 (see Figure \ref{fig6}.a). The curves seen in Figures \ref{fig6}.(a-b) were obtained by averaging the total energy or the absolute value of magnetization per each temperature in training and testing datasets (we labeled each microstate collected by ID-MH sampling by its energy, magnetization and temperature). Specifically,
\begin{align}
\langle | M | \rangle (T) &= \frac{1}{N^2} \sum_{i=1}^{i=40} \  \sum_{step =31,001}^{32,000}|M_{{seed}_i,\  {step}}(T)|,\\
\langle  E \rangle (T) &= \frac{1}{N^2} \sum_{i=1}^{i=40}\  \sum_{step =31,001}^{32,000} E_{{seed}_i,\  {step}}(T).
\end{align}
where \(T\) is a temperature variable taking values in the range from \(0.5\) to \(10\) with an increment \(\Delta T =0.5\) and \(step\) is a step of the ID-MH algorithm at which the current spin configuration is updated. 

The discrepancy in average training and testing values of magnetization and energy at low temperatures is due to the fact that the boundary energy states are excluded from the candidate testing subspace. Systems in the testing dataset do not get a chance to visit the lowest energy state. Overall, we conclude that the typical equilibrium behavior of the Ising-like ferromagnetic system is captured by training and testing data models created by the ID-MH sampling algorithm. 

\begin{figure}[!t]
\centering
\includegraphics[width=0.8\linewidth]{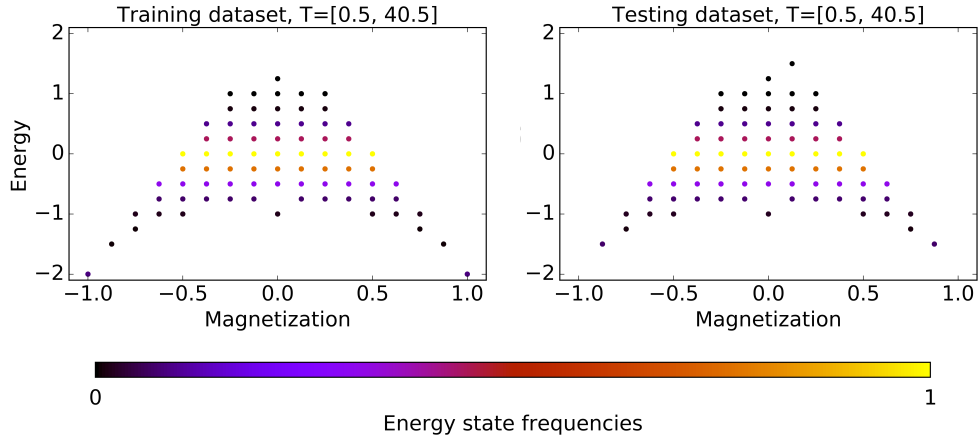}
\caption{Summary ``phase space" plots for the \(4 \times 4\) Ising model for the temperature range \([0.5, 40.5]\). The probability distributions of energy states are the same in the training and testing datasets created by ``ID-MH" algorithm.}
\label{fig7}
\end{figure}

Summary statistics of system visits of energy- and magnetization- dependent states over the temperature range \([ 0.5, 40.5]\) is shown by a color-coded graph in Figure \ref{fig7}. These training and testing datasets combine approximately canonical ensembles obtained for each temperature in the above mentioned range (with the temperature step equal to 0.5) by ID-MH sampling of the candidate training and testing subspaces. If we set the right hand side boundary temperature to a higher value, then we will be able to reach out for higher energy microstates. 

Here, we have illustrated the application of ID-MH for building the data models appropriate for learning equillibrium properties of the systems with a 2D lattice dimension \(N \leq 5\). However, if we are specifically interested in microstates representing higher energy states we can simply modify the temperature range by considering only high temperature values. Thus, the ID-MH algorithm provides with a flexible tool for building non-intersecting, consistent and physically motivated training and testing datasets depending on Ising-like system properties under study.       

\section{Block-ID sampling algorithm}
\label{sec3}
While ID-MH algorithm meets all requirements of the successful sampling strategy for validation of supervised learning algorithms for Ising-like systems, it appears to be of little practical interest. For example, the smallest repetitive volume that comprises the crystal pattern may contain \(88\) or more sites \cite{Guillaume:2011}. Application of ID-MH algorithm starts with the generation of the complete space of microstates leading to an astronomically large number of all possible configurations whose storage is impossible as it goes beyond the capacity of modern computers. In this section, we incorporate the ID-MH algorithm into our so-called \textit{block-ID} (bID) sampling strategy for large size Ising models, specifically, the ones whose size is a multiple of \(N \leq 5\). The bID sampling algorithm constitutes the central contribution of this paper.

\begin{figure}[ht]
\centering
\includegraphics[width=0.65\linewidth]{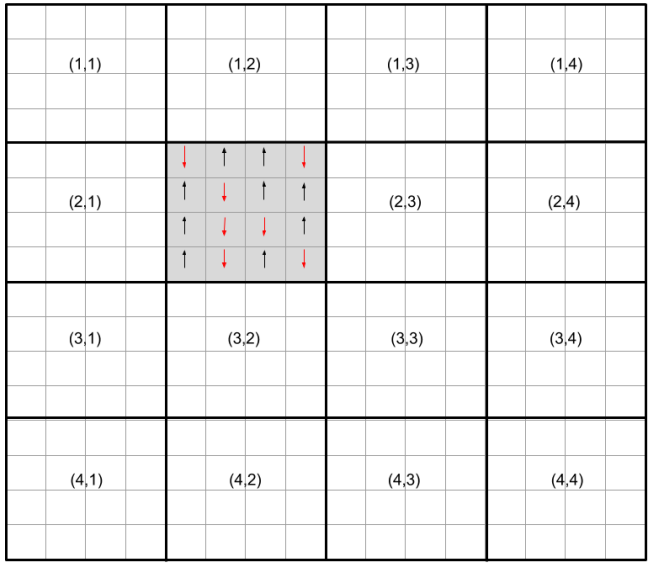}
\caption{Illustration of block-ID sampling algorithm for the \(16 \times 16\) Ising model viewed as a \(4 \times 4\) matrix of \(4 \times 4\) spin combinations. Shaded \(4 \times 4\) block represents a spin configuration randomly drawn from the candidate training or testing subspace. This block replaces the previous arrangement of spins at a site \((2,2)\), thus proposing an overall candidate \(16 \times 16\) configuration.} 
\label{fig8}
\end{figure}

We start with an observation that any configuration of spins with periodic boundary conditions arranged on a 2D lattice of size \((kN)^2\), where \(k \in \mathcal{N}\) and \(N \leq 5\) can be viewed as a composition of \(N \times N\) blocks consisting of \(1s\) and \(-1s\). More precisely, we can represent such configuration as a \(k \times k\) array of elementary \(N \times N\) blocks  (see Figure \ref{fig8}). In comparison to a single-spin-flip dynamics type of MC algorithm where we flip a spin at a random site of the current configuration to form a candidate one, we consider an \(N \times N\) block drawn from a randomly chosen energy deck of the candidate subspace as the proposed replacement at a random \( (i,j)\) site of the \(k \times k\) array.   
Given that Hamiltonian operator is defined with periodic boundary conditions where spins at the opposite boundary sites interact with each other, the Hamiltonian of the entire \(kN \times kN\) system is not the sum of Hamiltonians of its elementary \(N \times N\) blocks. As such, the difference in energy between the current and candidate systems denoted by \(C_l\) and \(C_n\), respectively, is computed globally, that is \(\Delta E=E(C_l)-E(C_n)\). With this approach, we have more choice of the candidates with lower than the current value of energy. Indeed, we can find \(kN \times kN\) candidates that will provide the energy difference of 8, 12, 16 or more units depending on the block dimension \(N\) (recall that with a classical single-spin-flip strategy, at a single MC step we can usually lower the energy by 4 units or by 8 units if transitioning to the lowest energy state). Also, keeping the acceptance rate as the one defined for MMC we immediately notice that it will be quite small for global candidates with energy values exceeding the current one by 8 units or more. As a result, the proposed bID technique described below step-by-step is more efficient than MMC as it ``walks" to an equillibrium distribution by larger MC steps. Moreover, it samples more variety of lower energy configurations. 
\bigskip
\newline
\textbf{\textit{Block-ID sampling algorithm.}}
\begin{enumerate}
\item {Choose a random configuration (initial seed) \(C_n\) of \(1s\) and \(-1s\) of size \((kN)^2\).}
\item{Structure \((kN)^2\) Ising system into \(k \times k\) array of \(N \times N\) blocks where \(N \leq 5\).}
\item{Generate candidate training and testing subspaces for \(N \times N\) Ising model using the splitting process of Section \textbf{\nameref{sec2}}.}
\item{Choose a random site \((i,j)\) of the block array of size \(k \times k\). }
\item{Draw a candidate energy deck \(m\) from the uniform distribution \(U({1,2,3,...,M})\), where \(M\) is the total number of energy states in the candidate subspace and \(m\) refers to one of the decks arranged in ascending order of their energy values.}
\item{Draw a random \(N \times N\) configuration from the candidate deck \(m\) and form a global candidate \(C_l\) of size \((kN)^2\) by replacing the elementary block at site \((i,j)\) with the selected configuration from the candidate subspace.}
\item {Compute the difference in energy \(\Delta E=E(C_l)-E(C_n)\).}
\item {Generate a random variable \(u \sim U(0,1)\)}.
\item {If \(\Delta E < 0\) or if \(\Delta E \geq 0\) and \(u < min(1, \exp{(-\beta \Delta E)}\) accept \(C_l\). Otherwise, stay at current configuration \(C_n\).}
\item {For each of the candidate subspaces, repeat steps 4 to 9 until convergence and record last 1000 global configurations along with their energies and magnetizations.} 
\end{enumerate}

We point out the advantages of uniform sampling over all energy decks/states in the candidate training or testing subspace. If we sample \(N \times N\) configurations at random then we will most likely select those ones in the vicinity of \(0th\) energy state (see Figure \ref{fig2}.a). If we sample energy states with \textit{ a priori} probability distribution of the complete space of \(N \times N\) configurations then the degeneracy of middle range energy states, will trap most of the candidates and slow convergence to the equillibrium distribution at low temperatures. If we sample energies according to the ID-MH algorithm maintaining the Markov chain across energy decks, then the energies of the accepted global Ising configurations will have a slow change comparable to that of MMC. By sampling uniformly all possible energy states we give an equal chance for them to occur including the lowest and highest energy states. For example, for \(N=4\) there is 1 chance out of 15 to choose the lowest energy state at each MC step of the stochastic process versus 2 chances out of 65,536 if the prior distribution is chosen (see Table \ref{tab1}). Thus, uniform sampling increases the chance to select less populated energy states and leads to a faster convergence of bID algorithm at low temperatures. In order to build consistent training and testing datasets, we perform block-ID sampling for the temperature range specified by a user. The reason for the temperature range will become apparent from the Section \textbf{\nameref{sec4}} below.  
\section{``Physics" and structure of block-ID sampled datasets}
\label{sec4}
\begin{figure}[!hb]
\centering
\includegraphics[width=0.8\linewidth, height=490pt]{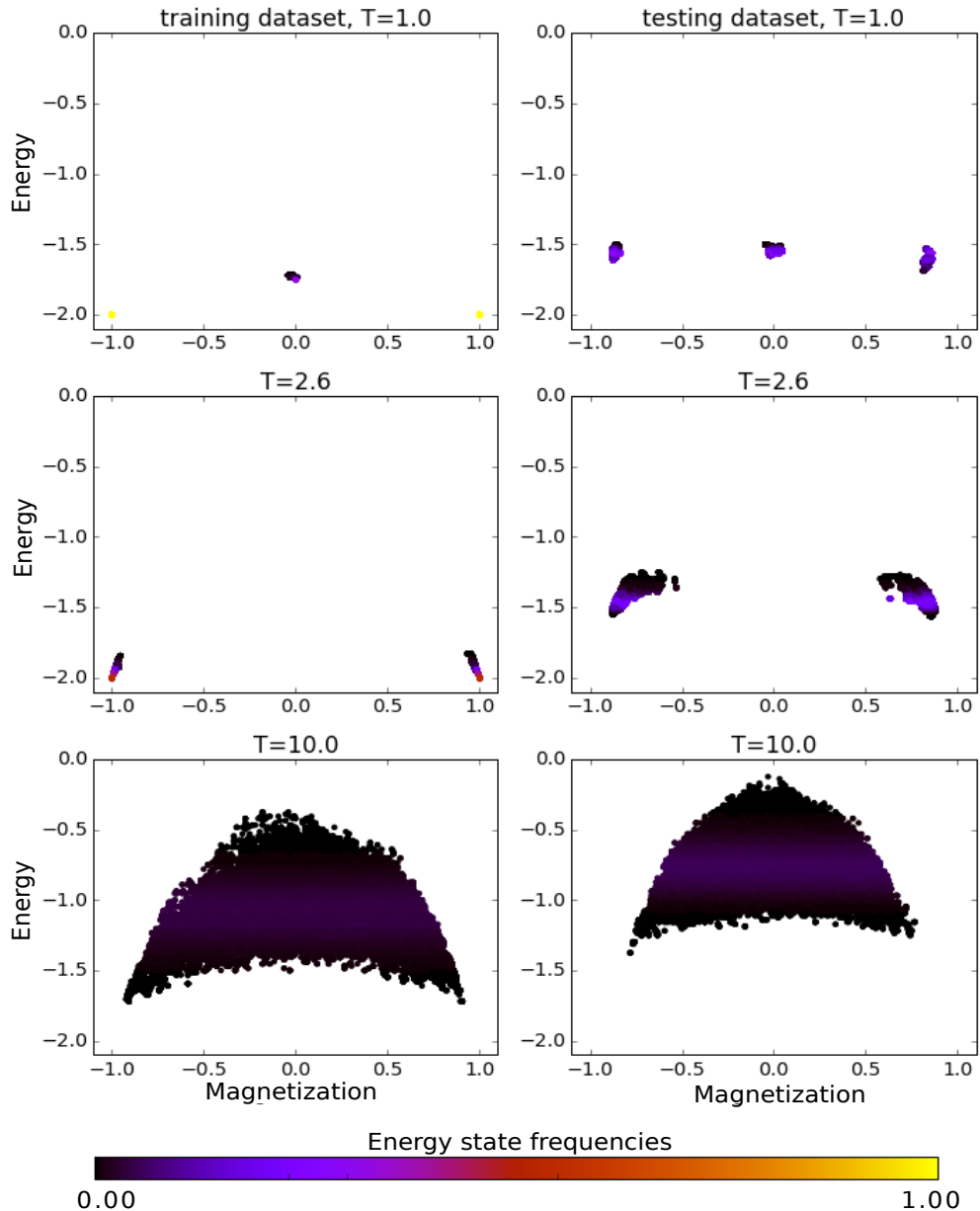}
\caption{``Phase space" plots of block-ID sampled training and testing data for the \(16 \times 16\) Ising model. An upward shift of visited energy states is observed in the testing set relative to the training set with the temperature decrease. Since the lowest energy state is excluded from the candidate testing subspace, at low temperatures the block-ID algorithm tends to select candidate configurations near the second lowest energy value for the \(4 \times 4\) Ising model.}
\label{fig9}
\end{figure}

We now run the bID sampling algorithm with 32,000 MC steps and 50 initial seeds, collecting 50,000 microstates in total for each given temperature \( \frac{1}{\beta}=1.0, 2.6, 10.0\) and plotting the frequency of visits of energy- and magnetization- dependent states for the \(16 \times 16\) Ising model (see Figure \ref{fig9}).
At low temperatures, most microstates in the training dataset tend to populate magnetization (per site) branches \( M =-1\) and \( M=1 \). The same behavior is observed in the testing data model, where the states with the lowest and highest net magnetizations present in the dataset are most frequently visited. Other configurations with zero magnetization (per site) are detected by the algorithm as they share the same low energy state. As temperature (\(\frac{1}{\beta}\)) increases to 2.6, the frequency of state visits extends to energy levels that neighbor with the lowest one. With respect to net magnetization per site, it follows magnetization branches starting at -1 and 1 in inward direction.  We can say that spontaneous magnetizations are maintained at low temperatures, consistent with statistical mechanics prediction of equillibrium behavior of ferromagnetic systems [\cite{Landau:2009}]. Increasing the temperature to 10.0 forms a parabolic triangular region of energies per site advancing to higher levels and magnetizations per site filling up the entire magnetization range. Unlike the case of \(4 \times 4\) Ising model, the frequency of states has not reached its peak at the \(0th\) energy level, settling in the lower energy range near the value of -1.0 in the training dataset, and near -0.75 in the testing dataset. 

\begin{figure}[ht]
\centering
\includegraphics[width=0.9\linewidth, height=210pt]{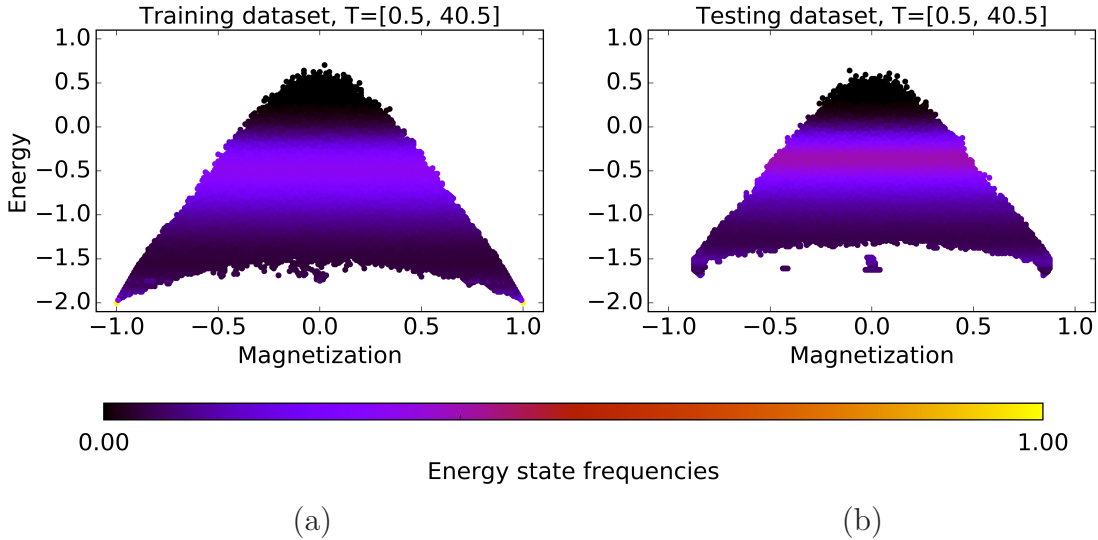}\\
\centerline{ \hspace{20pt} (a) \hspace{185pt} (b)}
\caption{Summary ``phase space" plots of block-ID sampled (a) training and (b) testing data for the \(16 \times 16\) Ising model. The training and testing datasets built by the block-ID algorithm are consistent as the testing energy range is included into the training energy range.}
\label{fig11}
\end{figure}

Comparison of energy ranges in training and testing datasets built by bID algorithm for each temperature shows an upward shift of the testing energy range (as a consequence of excluded boundary energy states) relative to the training one. This shift appears to be particularly pronounced for low temperatures when energy ranges in both datasets do not intersect. It will be impossible for a machine learning algorithm trained to recognize the energy range given in the training dataset to accurately predict other energy states that it has not learned. Training and testing datasets built per temperature are not consistent with respect to energy range and therefore, they cannot be used for algorithm validation. Training and testing data consistency is achieved by combining sets of microstates obtained by bID algorithm for different temperatures within a certain temperature range such as \([0.5, 40.5]\), for instance. Setting the temperature step to 0.5, we generated 80 sets of 50000 microstates labeled with their energy and magnetization values that we then combined into one dataset. Figure \ref{fig11} shows energy versus magnetization plots of such training and testing datasets where the states of the testing dataset form a subset of the state space of the training dataset. 

We also observe ferromagnetic phase transition from spontaneous magnetization occurring at the lowest energy level to its loss at higher positive energy levels. Given the tendency of bID algorithm to sample a variety of lower energy microstates the frequency peak is settled at the energy per site level between \(-0.5\) and \(0.0\).  

\begin{figure}[ht]
\centering
\includegraphics[width=.45\textwidth, height=160pt]{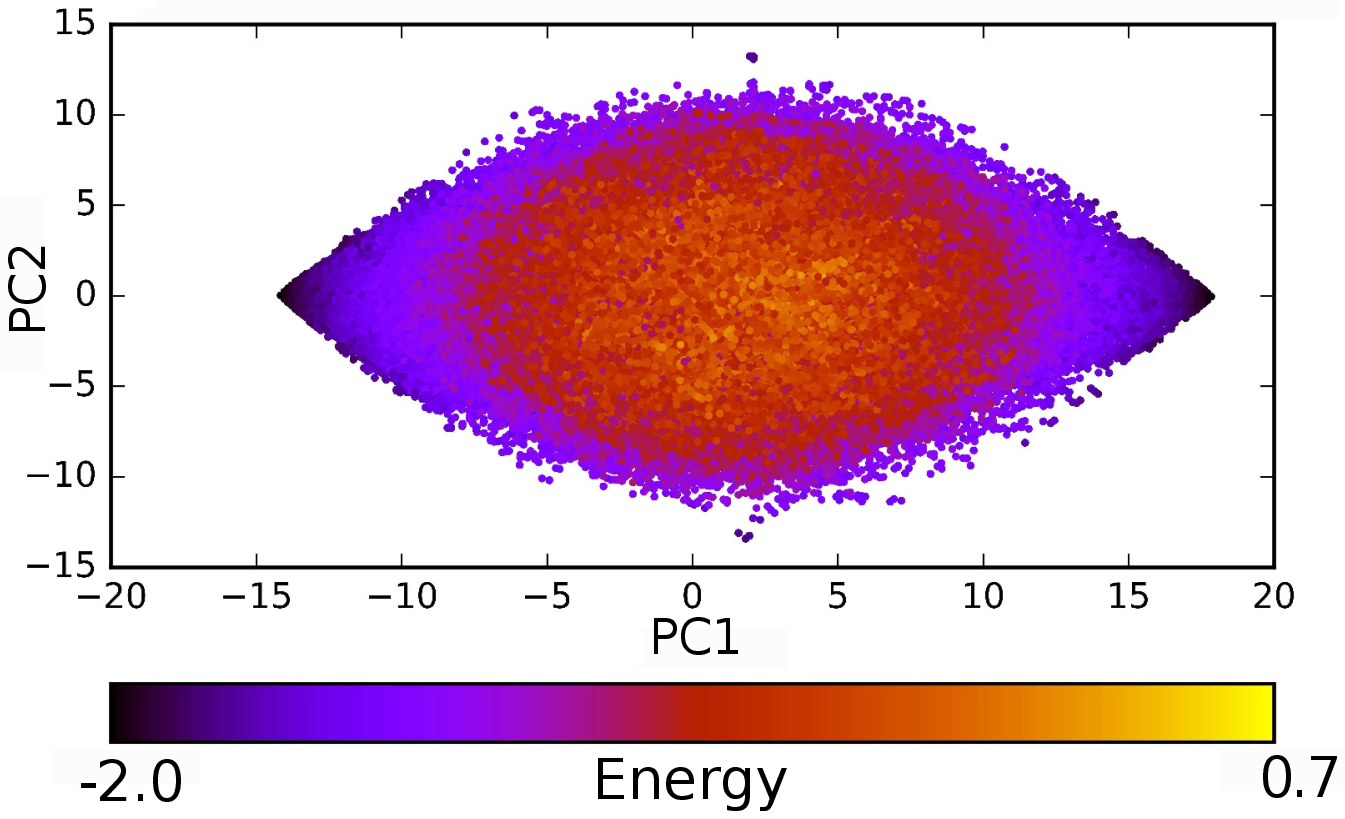}
\includegraphics[width=.45\textwidth, height=160pt]{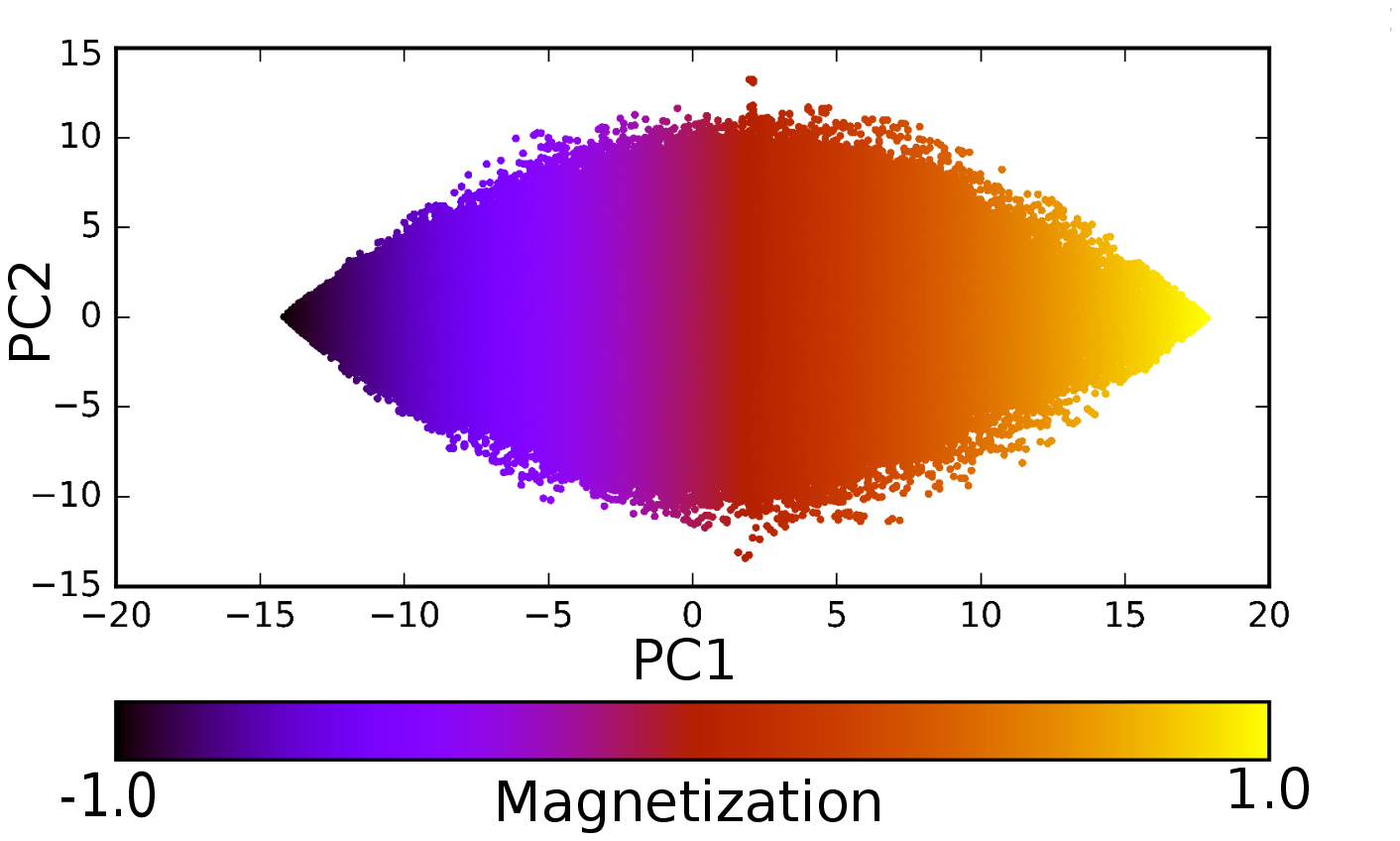}\\
\centerline{ (a) \hspace{180pt} (b)}
\caption{PCA view of the training dataset (4,000,000 samples) generated by block-ID sampling algorithm for the \(16 \times 16\) Ising model. Color-coded plots of (a) energies and (b) magnetizations.}
\label{fig12}
\end{figure}

We further explore the structure of bID sampled data, and since training and testing data models are alike, it suffices to consider the training dataset only. Having generated the data array \(X\) of size \(4,000,000 \times 256\) that contains 256 features (spin directions at 256 lattice sites) we perform Principal Component Analysis (PCA)-based dimensionality reduction that allows us to project the data onto the first two leading eigen-vectors \(\vec{e_1}\) and \(\vec{e_2}\) of the data covariance matrix \(X^T X\). Denote projected data coordinates by \(PC1\) and \(PC2\) in a new space spanned by  \(\vec{e_1}\) and \(\vec{e_2}\). Then for any observation \(\vec{X_k}\), where \(k \in \mathcal{N}:1 \leq k \leq 4,000,000\)
\begin{align}
PC1_k & =\langle \vec{X_k},{\vec{e_1}} \rangle,\\                           
PC2_k & =\langle \vec{X_k},{\vec{e_2}} \rangle. 
\end{align}
Figure \ref{fig12} shows a color-coded distribution of energies across the training dataset. Each colored point in this plot represents a microstate of the Ising system. The two lowest energy configurations with all spints down and all spins up correspond to the corners of the projected dataset and provide boundaries of the entire variability range along the X-axis. The ``hotter" the microstates are, the more similar (and closer to each other) they are. We can see the progression from less-alike microstates to more-alike in a structured manner, forming well-defined oval-shaped boundaries that distinguish between lower and higher energy configurations. There are only a few microstates with the positive energy shown in bright orange and yellow colors. The vast majority of configurations are within the energy range between \(-256.0\) and \(0.0\). They are displayed in shades of red. 

Figure \ref{fig12}.a shows that the dataset built by block-ID algorithm captures the symmetry of Hamiltonian operator. Indeed, a microstate with a positive x-coordinate in the 2D principal component space has a positive magnetization (see Figure \ref{fig12}.b) and its mirror image (with y-axis serving as a plane mirror) with a negative x-coordinate corresponding to the switching of the spin direction in all lattice sites of the microstate. The microstate and its mirror image possess the same energy. That is, \(\mathcal{H}(-x,y)=\mathcal{H}(x,y), \mathcal{H}(x,-y)=\mathcal{H}(x,y),\mathcal{H}(-x,-y)=\mathcal{H}(x,y)\).  Also, block-ID sampled data captures antisymmetric property of the magnetization function as clearly seen from Figure \ref{fig12}.b. Color-coded plot of magnetization values across the dataset shows that \(M(-x,y)=-M(x,y)\) holds for all configurations represented by (x,y) coordinates in the 2D principal component space (known as PCA scores).            

\section{Comparison of datasets obtained by MMC and block-ID methods}
\label{sec5}
\begin{figure}[!ht]
\begin{center}
    \begin{subfigure}[b]{0.45\textwidth}
        \includegraphics[width=\textwidth, height=160pt]{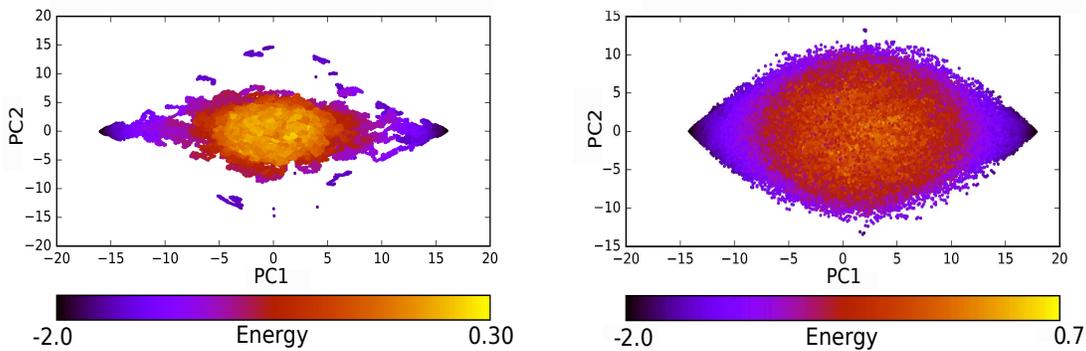}
        \centerline{(a)}
    \end{subfigure}
    %\hfill
    \begin{subfigure}[b]{0.45\textwidth}
        \includegraphics[width=\textwidth, height=160pt]{en_dep_pca_orig_block_IDv2.eps}
        \centerline{(b)}
    \end{subfigure}    
    \vspace{10pt}
    \begin{subfigure}[b]{0.8\textwidth}
        \includegraphics[width=\textwidth]{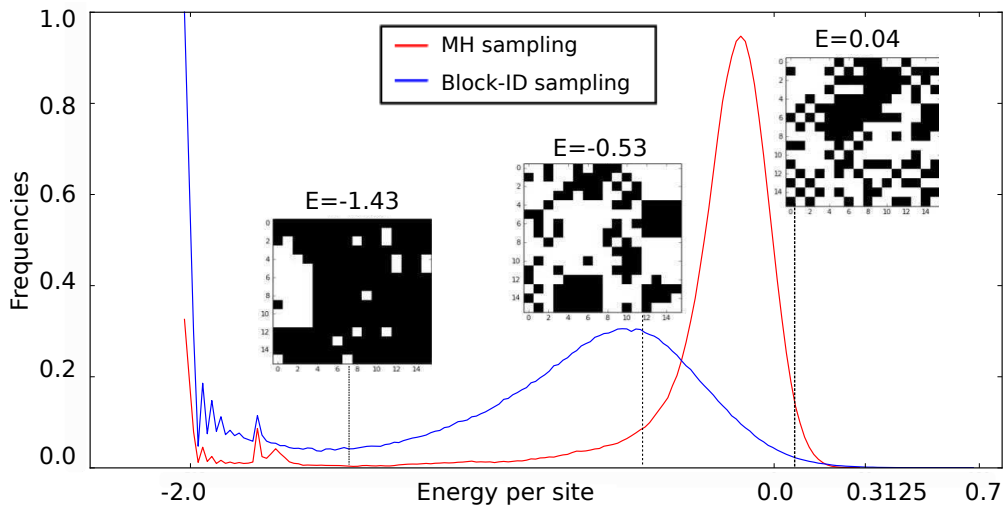}
        \centerline{(c)}
    \end{subfigure}    
\caption{(a-b) PCA view and (c) energy distributions of datasets generated by Metropolis-MC and block-ID sampling techniques for \(16 \times 16\) Ising model. Both datasets consist of 4,000,000 samples.}
\label{fig1}
\end{center}
\end{figure}

Having developed a new bID sampling strategy, we further justify it as our best choice for building supervised learning models of the Hamiltonian operator. For this purpose, we compare the structures of the datasets obtained by bID and MMC sampling for \(16 \times 16\) Ising model and shown in Figure \ref{fig1}.(a-b). 
Each dataset is comprised of 4,000,000 microstates collected at the asymptotic stage of the Markov chain. Namely, the global parameters for each algorithm were set to 50 initial seeds, the temperature range \([0.5, 40.0]\) with the sampling step of \( 0.5 \) and 32,000 MC steps (the total length of the Markov Chain). Upon reaching the equillibrium stage, the last one thousand configurations were recorded from each of the 50 Markov chains for every temperature in the given range. Dimensionality reduction of the datasets of size \(4,000,000 \times 256\) using Principal Component Analysis was then applied for visualization. It is clear that the MMC-generated dataset exhibits poor variability of the energy states(classes) in the energy range between dominant lowest energy and zeroth energy values. BID sampling compensates for the high imbalance of energy states by collecting more microstates from the underrepresented energy classes.

Figure \ref{fig1}.c compares the frequencies of visiting energy states collected by MMC and bID sampling algorithms. The bID algorithm undersamples the dominant zeroth energy class and adds microstates to the mid-range energy classes. At low temperatures, the Ising system stabilizes at the lowest energy state with a smaller probability to escape to a higher energy state (regulated by the acceptance rate) than in MMC asymptotics. As a result, thousands of repeated microstates form a peak at the lowest energy value. In reality, energies or other physical quantities characterizing a training molecular database are never evenly distributed, and therefore, we consider this data model as a true test of predictive capabilities of a supervised learning algorithm built on it. 

\begin{figure}[!ht]
    \centering
    \begin{subfigure}[b]{0.49\linewidth}
        \hspace{40pt}
        %\hfill
        \includegraphics[width=\linewidth]{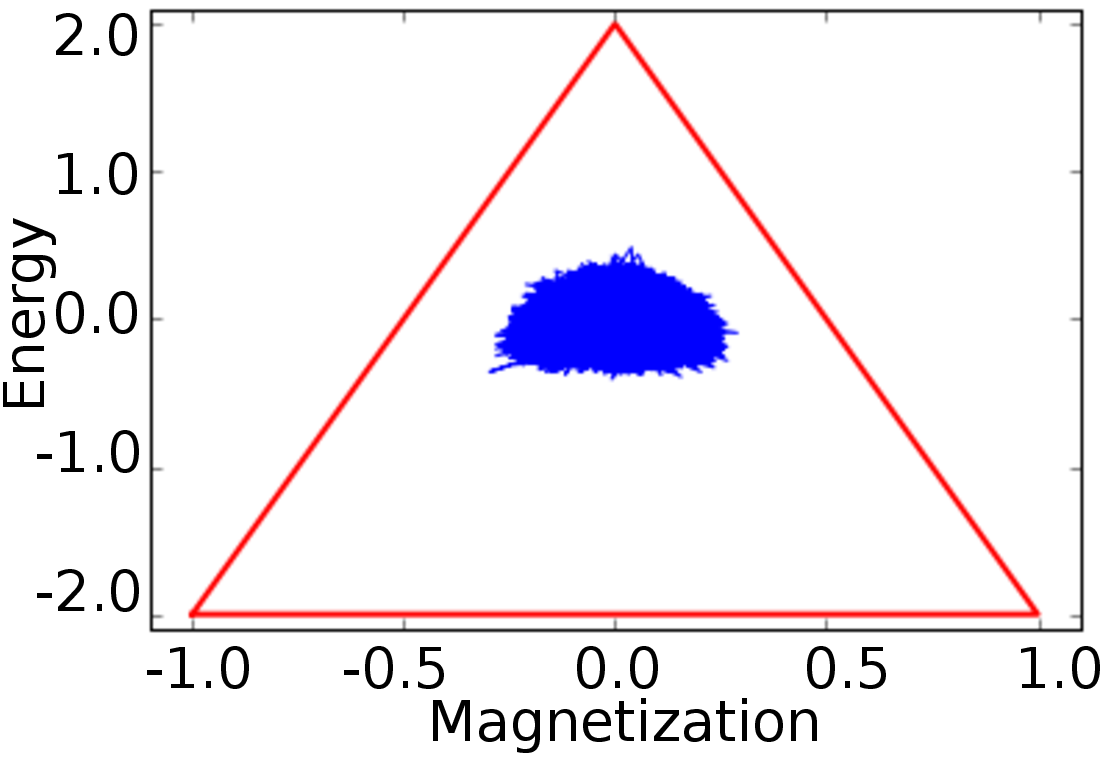} \hfill
        \centerline{\hspace{85pt} (a)}
    \end{subfigure}    
    \newline
    \begin{subfigure}[b]{0.45\textwidth}
        \includegraphics[width=\textwidth]{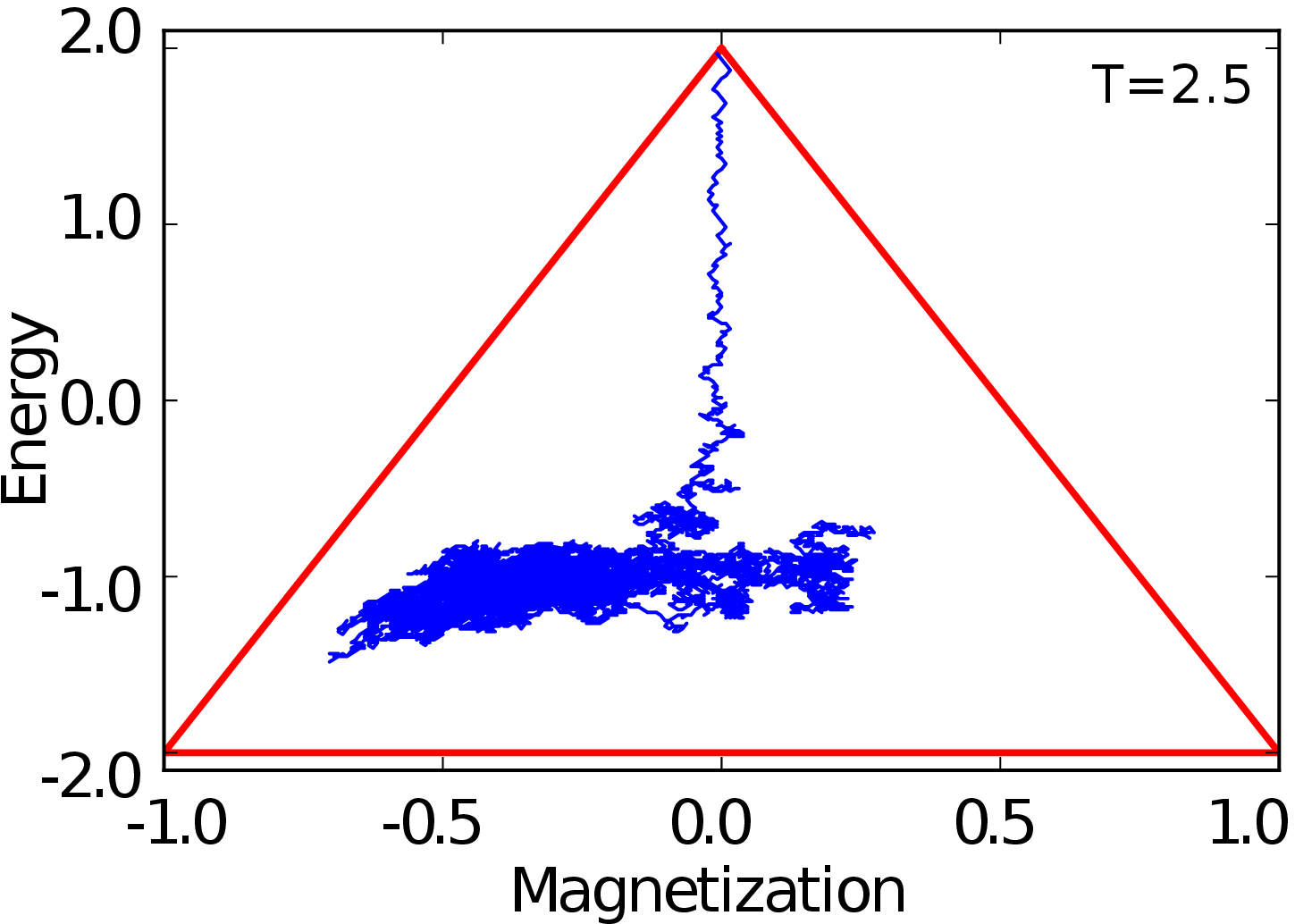}
        \centerline{(b)}
    \end{subfigure}
    \begin{subfigure}[b]{0.51\textwidth}
        \includegraphics[width=\textwidth]{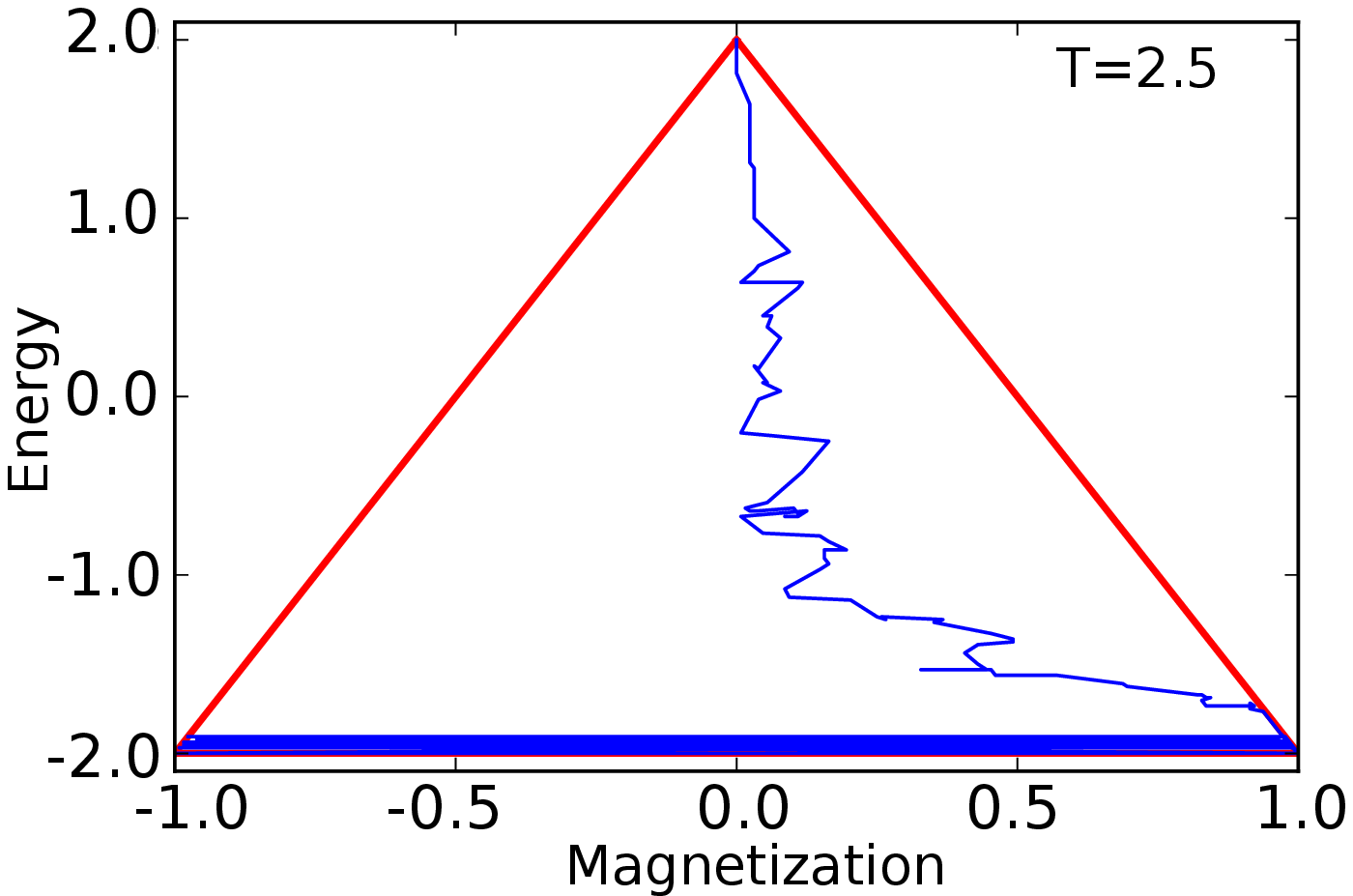}
        \centerline{(c)}
    \end{subfigure}    
\caption{Example seed trajectories obtained by (a) random, (b) Metropolis-MC and (c) bID sampling of \(16 \times 16\) Ising configurations for \(\frac{1}{\beta}=2.5\). Energies and magnetizations are computed per site (scaled by 256, the total number of lattice sites). The coupling coefficient \(J=1\).}
\label{fig2}
\end{figure}

When creating a dataset of microstates labeled with energies for learning a Hamiltonian operator, it is tempting to use to a simple random generator of Ising configurations as they can be represented by  \(N^2\)-dimensional vectors of \(1s\) and \(-1s\). As shown in Figure \ref{fig2}.a, a uniformly random sampling of about 4,000,000 spin configurations concentrates in the small neighbourhood of 0th energy value collecting a variety of disordered microstates. This is not surprising, since this energy class has the highest degeneracy that is, the highest number of microstate representatives. Given our interest in equillibrium properties of the Ising system, for temperatures below Curie's temperature, these microstates have very low Boltzmann weights and therefore, contribute very little to the actual value of physical observables. Random sampling does not capture microstates with important energies on the lower end of the range.

\begin{figure}[!h]
    \begin{center}
    \begin{subfigure}{.47\textwidth}
        \includegraphics[width=\textwidth]{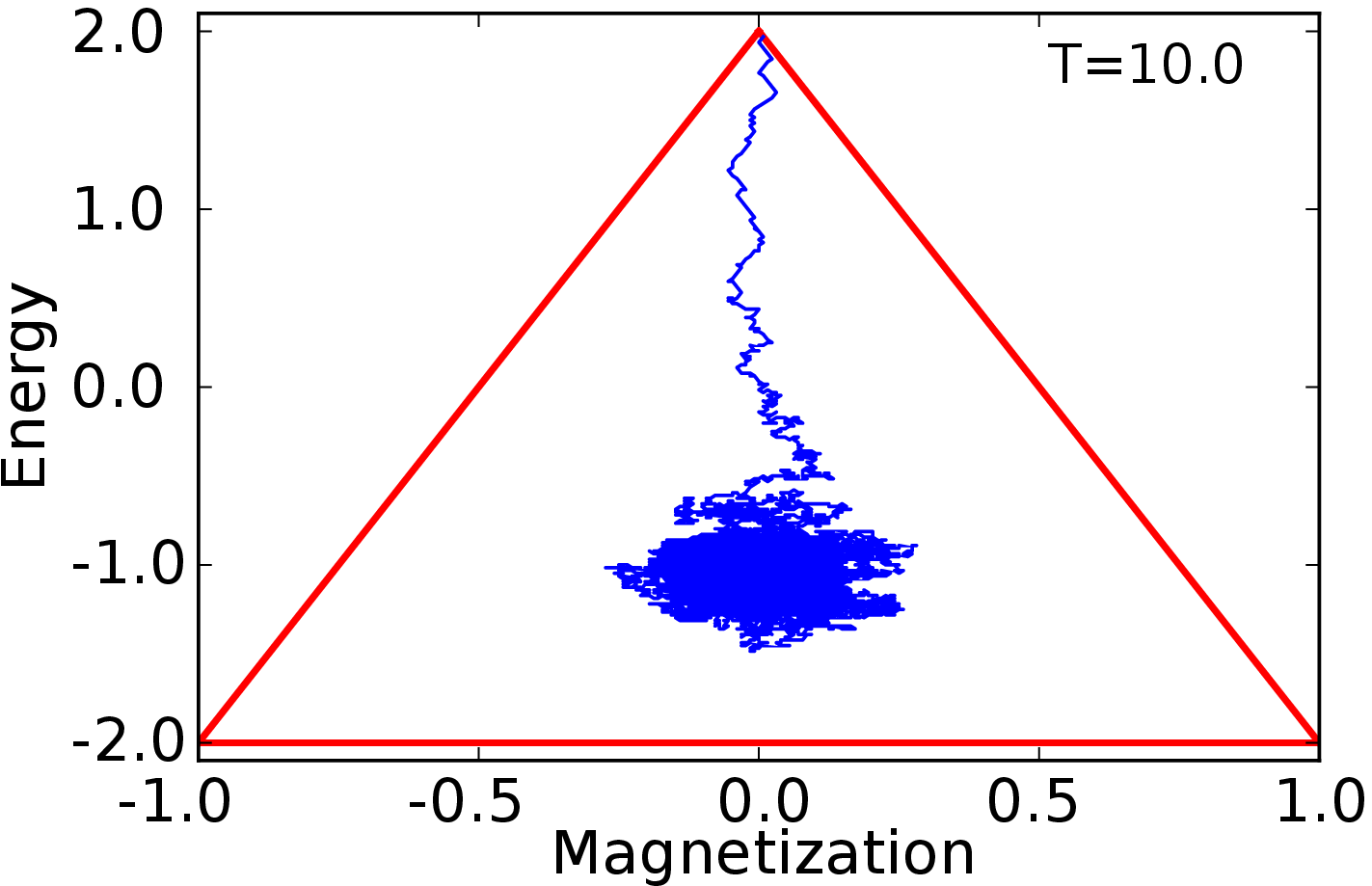}
        \centerline{(a)}
    \end{subfigure}
    \begin{subfigure}{.5\textwidth}
        \includegraphics[width=\textwidth]{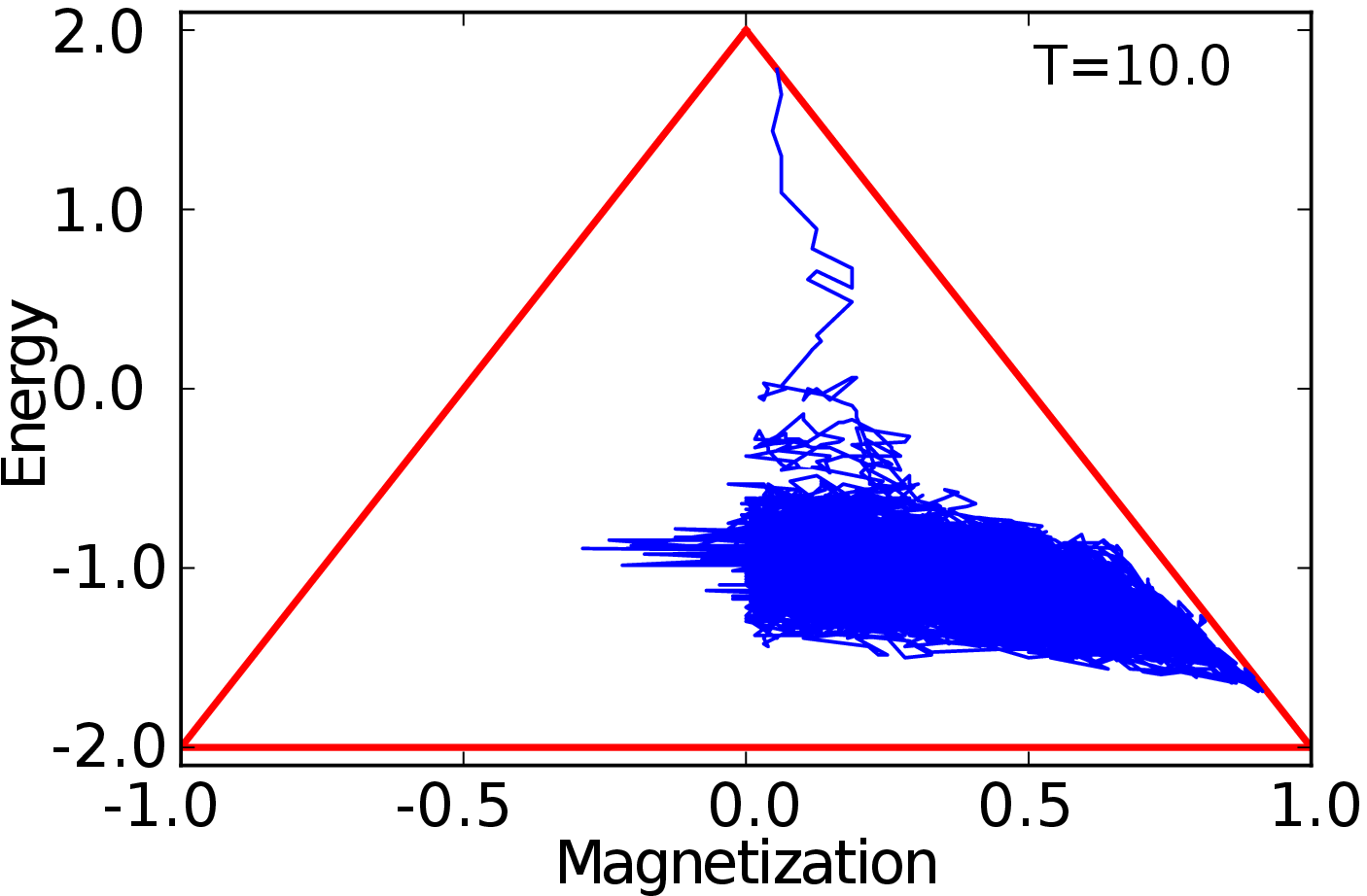}
        \centerline{(b)}
    \end{subfigure}    
\caption{Seed trajectories obtained by (a) Metropolis-MC and (b) bID sampling for \(16 \times 16\) Ising model for \(\frac{1}{\beta}=10.0\). Energies and magnetizations are computed per site. The coupling coefficient \(J=1\).}
\label{fig3}
    \end{center}
\end{figure}

Figures \ref{fig2}.b and \ref{fig3}.a show the difference between individual seed trajectories generated by MMC and bID sampling methods for \(\frac{1}{\beta}=2.5\) that start from the highest energy seed and proceed downwards visiting microstates in lower energy range. Total energies were computed with the coupling coefficient \(J=1\) for all pairs of interacting neighbors. BID sampling ``falls" to the lowest energy state oscillating between the two microstates with all spins oriented upwards and downwards. For \(\frac{1}{\beta}=10.0\) (see Figures \ref{fig2}.c and \ref{fig3}.b), the bID algorithm collects lower energy configurations with a significantly wider magnetization range as compared to Metropolis-MC.

Overall, the bID method gets to the lower energy range faster (MMC steps forming the trajectory are smaller as a consequence of this algorithm being a single-spin-flip dynamics type) and samples more variety of microstates within this range. 

\section{Machine learned operator performance comparison}
\label{sec6}
\subsection{Machine learning experiments with reduced size block-ID sampled datasets}
Having generated training and testing datasets containing 4,000,000 examples each via block-ID sampling, we explore different algorithmic models for learning Hamiltonian and Magnetization mappings. Given a 4 unit difference between the consecutive energy states and an 8 unit difference between the boundary states and their neighbors, we can formulate the energy forecast problem as the classification of Ising configuration to an energy state or class. Similarly, we can pose the classification problem for prediction of magnetizations since the difference between the consecutive values is 2 units. We can learn individual mappings using a variety of machine learning algorithms as well as two-label mappings that forecast energies and magnetizations simultaneously.

In this section, we intend to apply commonly used supervised learning methods such as \textit{decision trees (DT), random forests (RF), k nearest neighbors (kNN), artificial neural networks (ANN)} in both, classification and regression, paradigms. Each configuration for the \(16 \times 16\) Ising model is represented by a \(256\)-dimensional vector of \(1s\) and \(-1s\). In machine learning framework, the spins are referred to as features. As such, we have 4,000,000 observations of 256 features. As a fist simple and direct experiment to get an idea which paradigm suits best for our Ising data, we ran bID sampling algorithm to generate a smaller representative dataset with 5 initial seeds, 80 different values of temperature in the same range \([0.5, 40.5]\) of the ``global" dataset. For each seed trajectory, we only recorded the last 255 configurations, totaling in 102,000 examples overall. The testing dataset was generated in the same way.

\begin{figure}[!h]
\centering
\includegraphics[width=.45\textwidth, height=160pt]{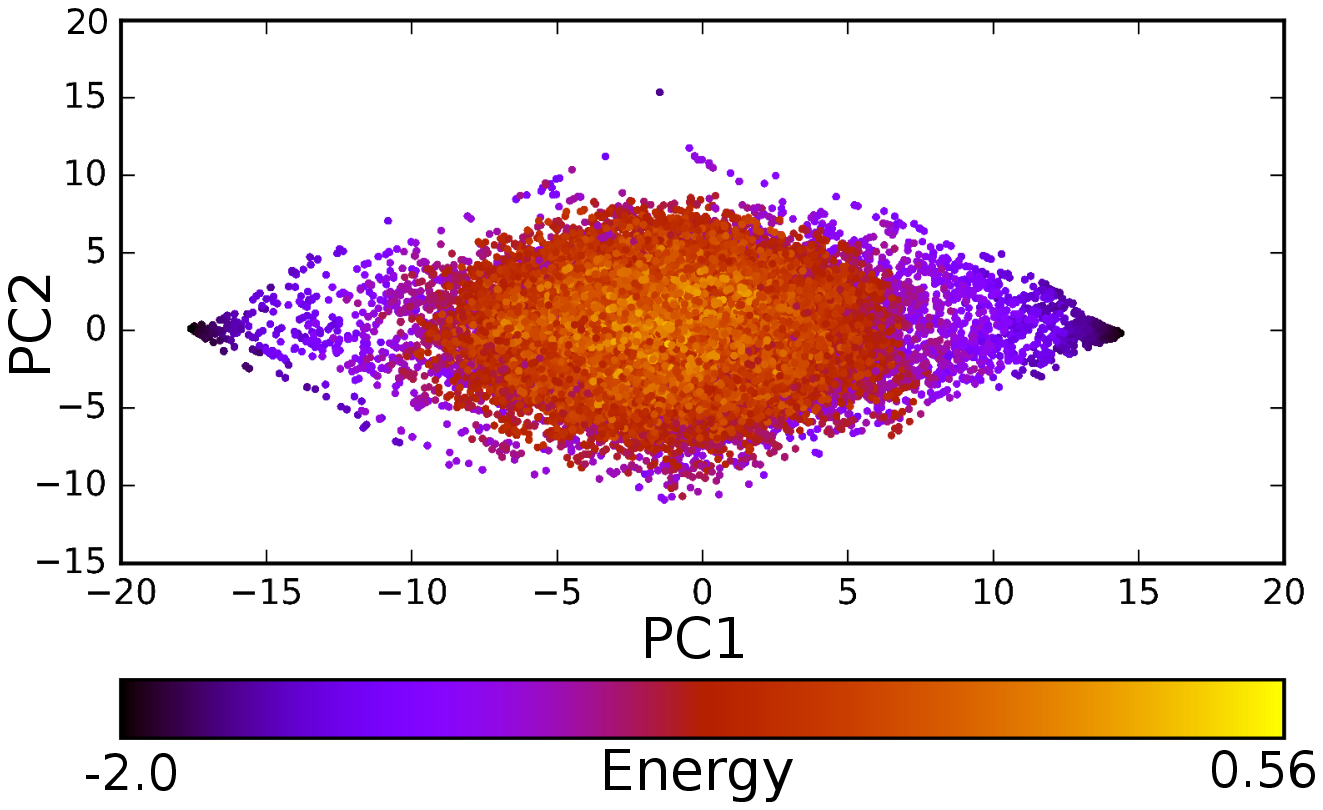}
\includegraphics[width=.45\textwidth, height=160pt]{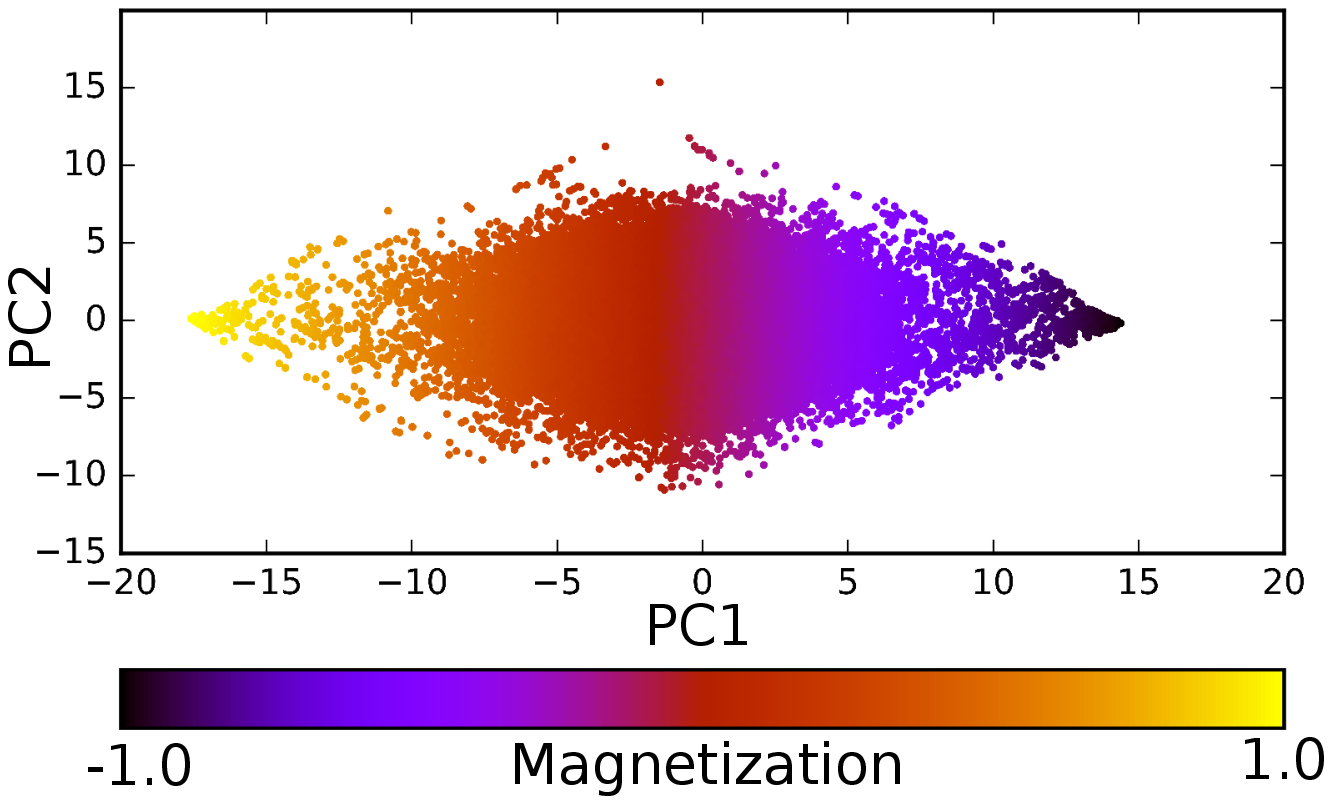}\\
\centerline{ (a) \hspace{160pt} (b)}
\caption{PCA view of the training subset (102,000 samples) generated by block-ID sampling algorithm for the \(16 \times 16\) Ising model. Color-coded plots of (a) energies and (b) magnetizations.}
\label{fig13}
\end{figure}

We then trained the above-mentioned classifiers and regressors individually for forecasting one target quantity, energy or magnetization, on raw Ising configurations of the reduced-size dataset shown in Figure \ref{fig13}. To access the performance of the trained models, we used a statistical measure of error (median). Since we have 102,000 examples in the testing dataset, for each example we computed the error as the absolute difference between the predicted and true values of energy or magnetization. We then computed the global median as the middle value of the sequence of 102,000 errors sorted in an ascending order. 

The first four rows of Table \ref{tab2} show the performance of DT, RF, kNN and ANN classifiers and regressors trained individually to predict energies and magnetizations. Given the \(\epsilon\)-distance of 4 units between neighboring energy classes, the median error appears unacceptably high for energies. ANN regressor is the best performer yielding a median error of 49.43. For magnetizations, ANN classifier yields the best accuracy results. In ANN calculations we used a simple architecture with one hidden layer and 256 hidden neurons. The median error of 0.0 means that in more than 50\% of testing samples we were 100\% correct in the magnetization prediction. Comparison of the error in the energy prediction yielded by classification and regression approaches shows that regression is best suited for the modeling Hamiltonian operator. kNN is the worst performer and as such it is discarded from further experiments with the global datasets. 

Another common machine learning strategy is data projection to a lower-dimensional space via Principal Component Analysis (PCA) and training a classifier/regressor on the projected data residing in this newly obtained space.  PCA is also viewed as a feature reduction technique compressing all important information about the variability of the dataset into a few linearly independent features. We calculate PCA features of the reduced-size dataset of size \(102,000 \times 256\) denoted by \(X_r\) as follows:
\begin{enumerate}
\item Compute the sample mean \(\vec{\mu}\).
\item Subtract the mean from each example in \(X_r\) and compute the data covariance matrix \((X_r-\vec{\mu})^T (X_r-\vec{\mu})\).
\item Compute the eigen-vectors and eigen-values of the covariance matrix and arrange eigen-vectors in the ascending order of their eigen-values.
\item Select the first few eigen-vectors \({\vec{e_1}, \vec{e_2}, .., \vec{e_L}}\) that have significant contributions to overall variance of the dataset (by examining their corresponding eigen-values).
\item Project the data sample (centered around \(\vec{0}th\) vector) onto the leading eigen-vectors selected in the previous step. If \(e\) is a matrix whose columns are the selected eigen-vectors, then \(Score=e^T (X_r-\vec{\mu})^T\). \(Score\) defines new coordinates of the projected data in PC space. 
\end{enumerate}
Using the above-described procedure, we extracted the 50 most important features (scores) that capture the variability of \(X_r\) in PC space. We then applied the same machine learning techniques to the PCA-based data representation. To prepare for prediction, testing examples were projected onto the 50 leading eigen-vectors of the training data covariance matrix. The error in energy prediction drops from 49.43 to 29.20 as seen from rows 5 to 8 of Table \ref{tab2}. PCA paired with Random Forests regressor appears to be the most accurate energy predictor. Overall, PCA transformation of Ising data significantly improves predictive capabilities of machine learning regressors lowering the error in energy prediction by at most 40\%. 

This is not surprising, as PCA highlights the data structure which is important to Decision Tree-based machine learning techniques. DT or RF select a hierarchy of features according to which the training dataset is cut into the subsets. These features are then extracted from an unknown example leading to the decision which subset it is most likely to belong to. As such, structured datasets are highly preferred when using these techniques as they are easier to subdivide.

The stripe-like structure of magnetization classes seen in Figure \ref{fig13}.b naturally suggests Decision Tree is a suitable algorithm. PCA paired with the Decision Tree regressor or classifier yields zero median error. Although ANN trained on raw Ising data also yields zero median error, PCA+DT is preferred due to its low computational cost of feature selection. We would like to carry over the best-performing algorithms to their training on massive datasets.

\begin{figure}[!h]
\centering
\includegraphics[width=0.99\linewidth]{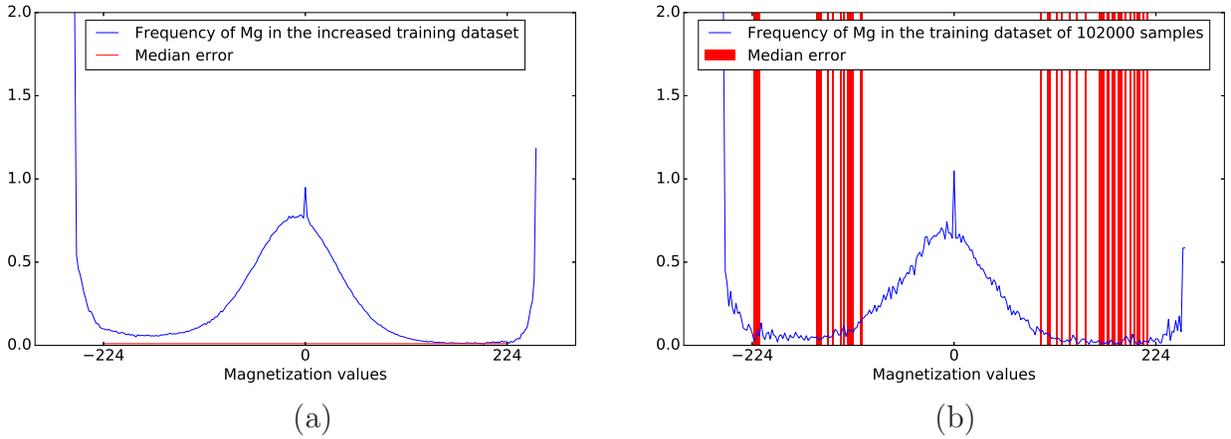}\\
\centerline{\hspace{10pt} (a) \hspace{220pt} (b)}
\caption{Median error of PCA-based Decision Tree regressor trained on (a) the dataset of 4,000,000 examples and (b) its subset of 102,000 examples. Increasing the size of the training dataset improves the accuracy of magnetization prediction. Original, non-uniform distribution of magnetizations in the training dataset does not affect the performance of the regressor.}
\label{fig14}
\end{figure}

Eventually, we formed training datasets with two labels per each example, one label for energy and another one for magnetization, and built a two-target predictor using PCA combined with DT and with RF to forecast energy and magnetization simultaneously. The errors for these predictors are reported in the last two rows of Table \ref{tab2}. As we can see from regression errors, simultaneous prediction of both quantities increases the median error in energy to 30.20 and in magnetization to 3.20. Therefore, regressors individually trained on datasets labeled with energies or datasets labeled with magnetizations constitute the best strategy for building supervised learning models of Hamiltonian and Magnetization operators.
\begin{table}[!ht]
\begin{tabularx}{0.9\textwidth}{|c|c|c|c|c|}
 \hhline{-----}
 ML method & Classification    & Regression & Classification & Regression \\
          & (\textbf{En})            & (\textbf{En})       & (\textbf{Mg})           & (\textbf{Mg})\\
 \hhline{-----}
 Decision Tree (DT)    & 80.00    & 72.00    & 40.00    & 36.00\\
 Random Forests (RF)    & 124.00    & 50.40    & 68.00    & 22.40\\
 K Near. Neighbors (KNN)    & 120.00    & 115.98    & 34.00    & 31.49\\
 A. Neural Networks (ANN) & 56.0 & 49.43 & 0.00 & 0.35 \\
 PCA features + DT    & 48.00    & 40.00    & 0.00    & 0.00 \cellcolor[gray]{0.7} \\
 PCA features + RF    & 92.00    & 29.20 \cellcolor[gray]{0.7} & 8.00    & 0.40\\
 PCA features + KNN    & 52.00    & 46.09    & 22.00    & 18.55\\
 PCA features + ANN    & 48.00    & 40.74    & 0.00    & 0.47\\
 Two-label, PCA + DT & 56.00    & 40.00    & 0.00    & 6.00\\
 Two-label, PCA + RF    & 96.00    & 30.80    & 4.00    & 3.20\\
 \hhline{-----}
\end{tabularx}
\caption{Global median errors in energy (\textbf{En}) and magnetization (\textbf{Mg}) prediction using the training dataset of 102,000 configurations generated by block-ID sampling. Lower is better.}
\label{tab2}
\end{table}
While the global median error reported in Table \ref{tab2} allows comparison of different machine learning strategies, it does not tell us about the distribution of the median error across energy or magnetization classes. The right hand-side plot of Figure \ref{fig14} shows the histogram distribution of the median error of PCA+DT regressor (with the global median error of \(0\)) across the entire range of magnetizations. The graph in blue color displays the frequencies of magnetization classes given in the reduced-size training dataset. We can now relate the error occurrences to the numbers of representatives per each magnetization value. As expected, small frequencies of magnetization class visits result in errors. However, these are still small and acceptable errors. The error of 2 units means that the magnetization class located next to the true one was forecasted.
\vspace{5mm}
 We can further improve the accuracy in magnetization prediction by increasing the size of the training dataset, specifically, by training on our global dataset of 4,000,000 examples. The left hand-side plot of the median error distribution shows the desired \(0th\) value across all magnetization classes even for seemingly underrepresented magnetization subranges. Since the testing dataset was created by the same bID sampling algorithm as the training one, the distribution of magnetizations in the testing dataset is similar to the one displayed by the blue graph in Figure \ref{fig14}. As such, it does not have a great variety of testing examples in the underrepresented magnetization subrange either resulting in \(0th\) median error.                   
\subsection{The effect of energy distribution equalization}
\begin{figure}[!ht]
\centering
\includegraphics[width=0.75\linewidth]{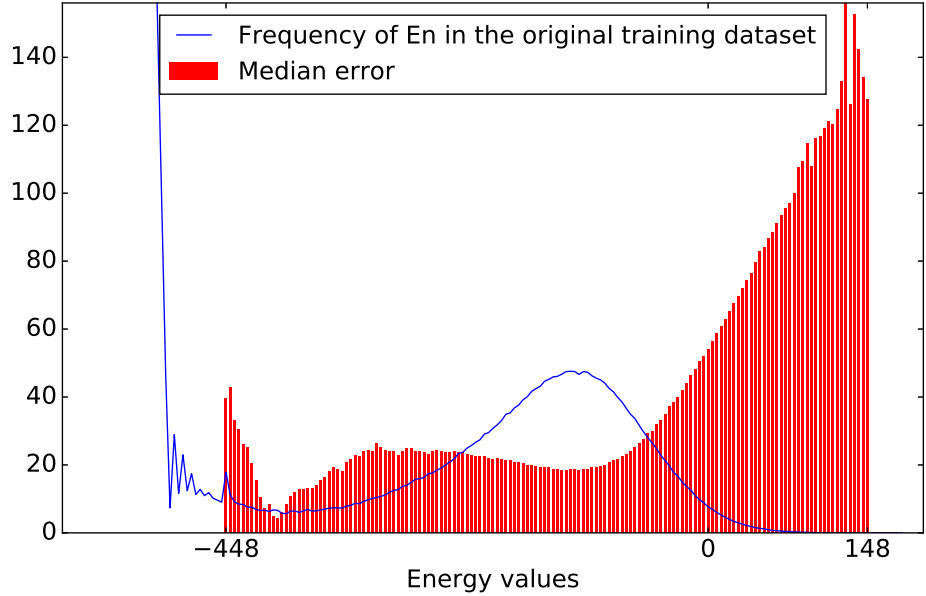}
\caption{Median error of PCA-based Random Forests regressor trained on original data of 4,000,000 samples (best performer for energies).}
\label{fig16}
\end{figure}

Having learned the best performing machine learning strategy from experiments presented in the above subsection, specifically PCA+RF regressor for prediction of energies, we now train this regressor on the global training dataset containing 4,000,000 examples. The first column of Table \ref{tab3} shows that this strategy remains the best performer among other methods trained on the same large size dataset. The global median error decreased in value by about 20\% due to the increased size of the training dataset. However, the error of 22.80 is still too big to be acceptable. The distribution of the median error across the energy classes seen in Figure \ref{fig16} for PCA+DT regressor shows high sensitivity of the predictor to underrepresented energy states. Indeed, since block-ID algorithm by construction mostly samples configurations on lower energy end, there are simply not enough positive energy samples in the training dataset for accurate prediction on higher energy end.        
As seen from Figure \ref{fig16} the major contribution to the global error comes from underrepresented positive energy states. 

In the experiment presented below, we studied the effect of equalization of energy distribution on the performance of supervised learning methods. To equalize the original histogram of energies, we computed the average number of representative samples per energy class (having excluded the dominant lowest energy class populated with repeated configurations). We set it to be the number of representatives per each energy class in the training dataset. For those classes with original numbers of representatives less than the average number we oversampled them with random copies of its representatives. Thus, having achieved the uniform distribution of energies we trained our usual regressors on the newly obtained dataset and compare the global median errors presented in Table \ref{tab3}. With respect to energies, the global error increased for all methods. For the forecast of magnetizations, equalization of energies caused no change in the error yielded by the best performing PCA+DT method. 

\begin{table}[!h]
\begin{tabularx}{0.9\textwidth}{ |c|c|c|c|c| }
 \hhline{-----}
 ML regression & Original & Equalized & Original & Equalized \\
 method & distr. (\textbf{En}) & distr. (\textbf{En}) & distr. (\textbf{Mg}) & distr. (\textbf{Mg})\\
 \hhline{-----}
 DT    & 68.00    & 68.00    & 32.00    & 32.00\\
 RF    & 45.20    & 46.80    & 19.40    & 20.80\\
 ANN    & 42.84    & 48.81    & 0.12    & 0.12\\
 PCA features + DT    & 32.00    & 36.00    & 0.00 \cellcolor[gray]{0.7}    & 0.00\\
 PCA features + RF    & 22.80 \cellcolor[gray]{0.7}    & 24.40    & 0.00    & 0.00\\
 PCA features + ANN    & 36.73    & 41.15    & 0.18    & 0.28\\
 Two-label, PCA + DT & 32.00    & 36.00    & 4.00    & 4.00\\
 Two-label, PCA + RF    & 23.60    & 25.20    & 2.60    & 3.20\\
 \hhline{-----}
\end{tabularx}
\caption{Global median errors in energy (\textbf{En}) and magnetization (\textbf{Mg}) prediction using the training dataset of increased size (4,000,000 samples) and its modified version with equalized distribution of energies.}
\label{tab3}
\end{table}

\begin{figure}[!h]
\centering
\includegraphics[width=0.99\linewidth]{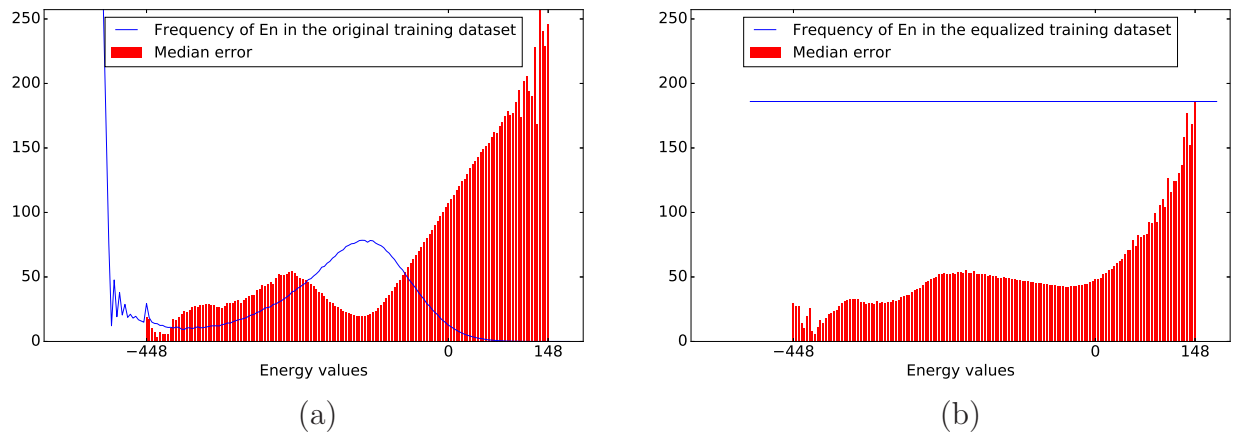}\\
\centerline{\hspace{10pt} (a) \hspace{220pt} (b)}
\caption{Median error of PCA-based Artificial Neural Networks regressor trained on data with (a) original and (b) equalized energy distributions. Equalization of energy distribution in the training dataset lowers the error of prediction of underrepresented `tail` energy classes.}
\label{fig15}
\end{figure}

How does equalization of energy distribution impact the histogram of median errors across the energy range? Figure \ref{fig15} shows the median error histogram produced by PCA-based ANN regressor trained on the global dataset with original (see Figure \ref{fig15}.a) and its modified version with equalized distribution of energies (see Figure \ref{fig15}.b). As a result, the error distribution tends to equalize across the energy range. The error drops significantly over the subrange of positive energies while raising in the middle energy subrange with a maximal number of original representative microstates. Apparently, adding repeated samples to minority energy classes improves accuracy of prediction for those energies. This is an important finding that we can use, for instance, to improve forecasting of energy classes that are naturally underrepresented for Ising systems. Specifically, these are lowest and highest energies that appear important in calculating equillibrium and non-equillibrium properties of the system, respectively. For instance, the lowest energy class appears to have the biggest Boltzmann weight for systems at thermal equillibrium at low temperatures. In this case, training on the Ising dataset with equalized energy distribution is preferable, as other energy classes such as the ones in the middle energy range will have a small contribution to computing physical observables of the system. 
Therefore, training datasets can be adjusted to target certain energy classes that we know \textit{a priori} are important to predict accurately.
\newline

\section{Conclusion}
\label{sec7}
In this paper, we illuminated subtle drawbacks of classical Metropolis-MC sampling methods for the Ising model that prevent the use of conventional validation approaches for supervised learning models of the Hamiltonian operator. We developed new MC-based methods that overcome a sampling bias inherent to Metropolis-MC and serve as a validation tool for machine learning algorithms. The proposed sampling methods (ID-MH algorithm for lattice dimension \(N \leq 5\) and block-ID algorithm for lattice dimension equal to the multiple of \(N\)) accomplish the following tasks:
\begin{enumerate}
\item Sample a greater variability of Ising configurations that represent important energy states of the system at a thermal equilibrium,  
\item Build distinct training and testing datasets needed for a fair assessment of machine learning algorithm performance,
\item Build physically motivated training and testing data models that capture typical equillibrium behavior, specifically, a phase transition for ferromagnetic systems,
\item Collect a structured ensemble of microstates that captures the symmetry of Hamiltonian operator and antisymmetry of Magnetization operator, 
\item Ensure consistency between training and testing datasets that avoids mismatch of energy states in the testing and training sets and avoids imbalance between numbers of representatives of the same energy state present in the training and testing sets.
\end{enumerate}
We shared important insights into the error behavior gained from machine learning experiments with block-ID sampled datasets for the 2D Ising model. The global median error is not informative and therefore, we computed the histogram distribution of the median error across all energies and all magnetizations of the testing dataset. The median error distribution for energies revealed a high sensitivity of the best performing machine learning strategy to underrepresented positive energy states in the training dataset. The median error distribution for magnetizations did not react to low-frequency magnetizations present in the training data and was found to be equal to \(0\). The best performing method for prediction of magnetizations is PCA-based Decision Tree regression yielding the median error of \(0\) per each magnetization value in the testing set. 

For prediction of energies, our findings show that regression approach works best and PCA-based data representation significantly improves the accuracy of commonly used algorithms (DT, RF, kNN, ANN). The winner among those algorithms is PCA-based Random Forests regressor trained on a large size block-ID sampled dataset of 4,000,000 Ising configurations. However, it yields the median error of 22.80 units that is unsatisfactory given the \(\epsilon\)-distance of 4 units between two consecutive energies of the Ising system. Therefore, the best performer for forecasting energies is yet to be found. Exploration of deep learning methods with predefined architectures that have been found successful in various object recognition applications will be our next research endeavour.          

\section{Acknowledgement}
The authors acknowledge NSERC, SOSCIP and Compute Canada for funding and computational resources.

\bibliography{sample}

\end{document}